%% file: HIG-20-014_temp.tex
\pdfoutput=1
\documentclass[11pt,twoside,a4paper,cmspaper,final,collab]{cms-tdr}
\def\svnVersion{ecf23bb}\def\svnDate{2021/11/15}\def\cmsCernNoTag{CERN-EP-2021-094}\def\cmsCernDate{\today}\def\cmsMessage{Published in the Journal of High Energy Physics as \href{http://dx.doi.org/10.1007/JHEP11(2021)057}{\doi{10.1007/JHEP11(2021)057}.}}
\begin{document}\cmsNoteHeader{HIG-20-014}

\newcommand{\PS}{\ensuremath{\text{S}}\xspace}
\newcommand{\Phs}{\ensuremath{\Ph_{\text{S}}}\xspace}
\newcommand{\Irelem}{\ensuremath{I_{\text{rel}}^{\Pe(\mu)}}\xspace}
\newcommand{\ptem}{\ensuremath{\pt^{\Pe(\Pgm)}}\xspace}
\newcommand{\mtautau}{\ensuremath{m_{\Pgt\Pgt}}\xspace}
\newcommand{\mbb}{\ensuremath{m_{\PQb\PQb}}\xspace}
\newcommand{\emu}{\Pe\Pgm}
\newcommand{\mutau}{\Pgm\tauh}
\newcommand{\etau}{\Pe\tauh}
\newcommand{\tautau}{\tauh\tauh}
\newcommand{\FF}{\ensuremath{F_{\text{F}}}}
\newcommand{\jettau}{\ensuremath{\text{jet}\to\tauh}}
\newcommand{\Jettau}{\ensuremath{\text{Jet}\to\tauh}}
\newcommand{\ZEE}{\mbox{\ensuremath{\PZ\to\Pe\Pe}\xspace}}
\newcommand{\ZMM}{\mbox{\ensuremath{\PZ\to\Pgm\Pgm}\xspace}}
\newcommand{\Wjets}{\ensuremath{\PW}{+}\text{jets}\xspace}
\newcommand{\xsecDotBR}{\ensuremath{\sigma\,\mathcal{B}(\PH\to\Ph(\Pgt\Pgt)\Phs(\PQb\PQb))}}
\newcommand{\yDT}{\ensuremath{y^{\text{DT}}}}
\newcommand{\mH}{\ensuremath{m_{\PH}}}
\newcommand{\mh}{\ensuremath{m_{\Ph}}}
\newcommand{\mhs}{\ensuremath{m_{\Phs}}}
\newcommand{\Dj}{\ensuremath{D_{\text{jet}}}}
\newcommand{\De}{\ensuremath{D_{\Pe}}}
\newcommand{\Dm}{\ensuremath{D_{\Pgm}}}
\newcommand{\ttc}{\ensuremath{\text{tt}}}
\newlength\cmsTabSkip\setlength{\cmsTabSkip}{2ex}

\cmsNoteHeader{HIG-20-014}
\title{Search for a heavy Higgs boson decaying into two lighter Higgs bosons in the \texorpdfstring{$\Pgt\Pgt\PQb\PQb$}{tautaubb} final state at \texorpdfstring{$13\TeV$}{13 TeV}}

\date{\today}

\abstract{
   A search for a heavy Higgs boson $\PH$ decaying into the observed Higgs boson $\Ph$ with a mass of $125\GeV$ and another Higgs boson $\Phs$ is presented. The $\Ph$ and $\Phs$ bosons are required to decay into a pair of tau leptons and a pair of $\PQb$ quarks, respectively. The search uses a sample of proton-proton collisions collected with the CMS detector at a center-of-mass energy of $13\TeV$, corresponding to an integrated luminosity of $137\fbinv$. Mass ranges of $240$--$3000\GeV$ for $\mH$ and $60$--$2800\GeV$ for $\mhs$ are explored in the search. No signal has been observed. Model independent $95\%$ confidence level upper limits on the product of the production cross section and the branching fractions of the signal process are set with a sensitivity ranging from $125\unit{fb}$ (for $\mH=240\GeV$) to $2.7\unit{fb}$ (for $\mH=1000\GeV$). These limits are compared to maximally allowed products of the production cross section and the branching fractions of the signal process in the next-to-minimal supersymmetric extension of the standard model.
}

\hypersetup{%
pdfauthor={CMS Collaboration},%
pdftitle={Search for a heavy Higgs boson decaying into two lighter Higgs bosons in the tautau bb final state at 13 TeV},%
pdfsubject={CMS},%
pdfkeywords={CMS, higgs, BSM, NMSSM}}

\maketitle 
\section{Introduction}
\label{sec:introduction}

The discovery of the Higgs boson ($\Ph$) with a mass of $125\GeV$ at the CERN 
LHC~\cite{Aad:2012tfa,Chatrchyan:2012xdj,Chatrchyan:2013lba} has turned the 
standard model (SM) of particle physics into a theory that could be valid 
up to the Planck scale. To date all properties of the observed particle are in 
agreement with the expectations of the SM within an experimental precision of 
$5$--$20\%$~\cite{Khachatryan:2016vau,Sirunyan:2018koj,Aad:2019mbh,
Sirunyan:2019twz}. Despite its success in describing a wealth of phenomena, 
the SM falls short of addressing a number of fundamental theoretical questions 
and striking observations in nature. In this respect it is considered to be 
still incomplete.

Supersymmetry (SUSY) postulates a bosonic (fermionic) partner particle for each 
SM fermion (boson), with the same quantum numbers as the corresponding SM particle 
apart from its \mbox{(half-)} integer spin~\cite{Golfand:1971iw,Wess:1974tw}. 
The fact that to date no such SUSY particles have been observed implies that if 
SUSY were realized in nature it must be a broken symmetry. Apart from the prediction 
of a sizable number of new particles, SUSY requires the extension of the 
Brout-Englert-Higgs mechanism part~\cite{Higgs:1964ia,Higgs:1964pj,Guralnik:1964eu,
Englert:1964et,Higgs:1966ev,Kibble:1967sv} of the SM Lagrangian. In the 
minimal supersymmetric extension of the SM (MSSM)~\cite{Fayet:1974pd,Fayet:1977yc} 
one more $\text{SU}(2)$ doublet of complex scalar fields is introduced with 
respect to the SM, leading to the prediction of two charged and three neutral 
Higgs bosons, one of which can be associated with $\Ph$. A further extension of 
the MSSM by one additional complex scalar field $\PS$ is theoretically well 
motivated, since it can solve the so called ``$\mu$-problem'' of the 
MSSM~\cite{Kim:1983dt}. It leads to the next-to-minimal supersymmetric SM (NMSSM), 
as reviewed, \eg, in Refs.~\cite{Ellwanger:2009dp,Maniatis:2009re}. Since $\PS$ 
is a complex field, the number of predicted Higgs bosons increases by two, 
resulting in two charged and five neutral Higgs bosons, of which three are scalar 
and two are pseudoscalar in nature.   

Many searches for additional Higgs bosons in the context of the MSSM have been 
performed by the LHC experiments. In the absence of signal, these have led to 
the exclusion of large parts of the MSSM parameter space for masses of the 
additional neutral Higgs bosons up to ${\approx}2\TeV$~\cite{Aad:2020zxo,
Sirunyan:2018zut,Aaboud:2018gjj,Sirunyan:2019hkq}. The parameter space of the 
NMSSM, on the other hand, is still largely unconstrained~\cite{King:2014xwa}. 

The current analysis focuses on the $\PH\to\Ph\Phs$ decay of a heavy Higgs boson 
$\PH$ into $\Ph$ and another neutral boson $\Phs$ with a mass of $\mhs<\mH-\mh$. 
It is based on the data recorded during the years 2016, 2017, and 2018 at a 
center-of-mass energy of $13\TeV$ with the CMS experiment, resulting in an 
integrated luminosity of $137\fbinv$. The search is inspired by the NMSSM, where 
$\Phs$ could have a dominant admixture of the additional singlet field $\PS$, 
leading to a significant suppression of its couplings to SM particles and thus 
of its direct production at the LHC. In this case, the production of $\PH$ 
and subsequent decay into $\Ph\Phs$ would become the dominant source for $\Phs$ 
production. Despite the overall reduced coupling strengths to SM particles, the 
branching fractions of $\Phs$ for its decay into SM particles are still expected 
to be similar to those of $\Ph$. While here we use the NMSSM as a motivation, 
any other two Higgs doublet plus singlet model is equally relevant for the 
search.

\begin{figure}[t]
  \begin{center}
    \includegraphics[width=0.6\textwidth]{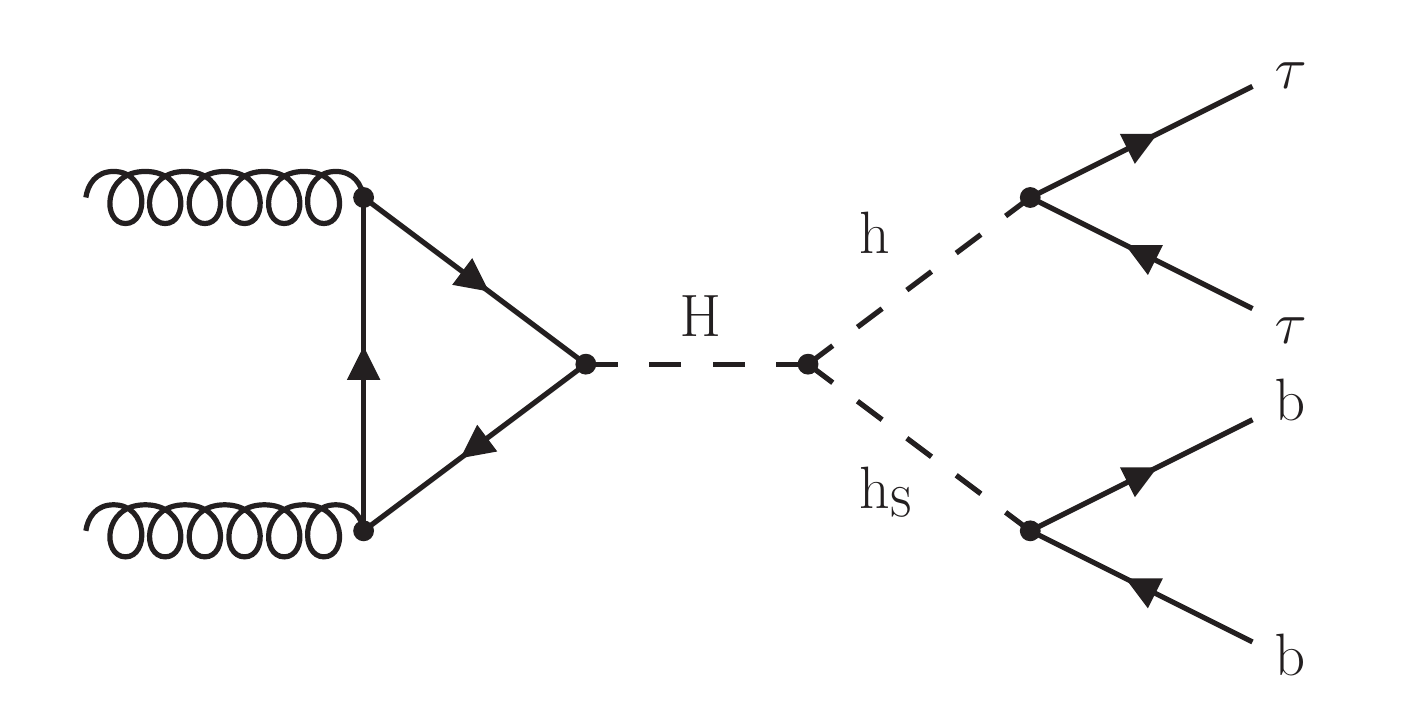}
    \caption{
      \label{fig:process}Feynman diagram of the $\Pg\Pg\to\PH\to\Ph(\Pgt
      \Pgt)\Phs(\PQb\PQb)$ process.
    }
  \end{center}
\end{figure}

A promising signature for the search is given by the decay of $\Ph$ into a 
pair of tau leptons and the decay of $\Phs$ into a pair of $\PQb$ quarks, 
$\Ph(\Pgt\Pgt)\Phs(\PQb\PQb)$. For better readability we will not distinguish 
fermions by particle or antiparticle in this final state in subsequent notation 
throughout the text. The decay into $\PQb$ quarks is chosen for its large branching 
fraction. The decay into tau leptons is chosen for its cleaner signature compared 
to the decay into $\PQb$ quarks. This search is restricted to $\PH$ production 
from gluon fusion. The Feynman diagram for the process of interest is shown in 
Fig.~\ref{fig:process}. The search is performed in the mass ranges of $240\leq\mH
\leq 3000\GeV$ and $60\leq\mhs\leq 2800\GeV$. It is the first search for such a 
process at the LHC. No attempt is made to identify and treat specially boosted 
topologies, for which the $\Ph$ and $\Phs$ decay products may not easily be 
spatially resolved. These can occur in parts of the explored mass ranges, \eg, 
for large values of $\mH$ and small values of $\mhs$. However, for the majority 
of the mass hypotheses that are considered, the contribution from boosted-topology 
events is subdominant.

The paper is organized as follows. A brief introduction of the CMS detector and 
event reconstruction are given in Sections~\ref{sec:detector} 
and~\ref{sec:event-reconstruction}, respectively. The model used to describe 
the data is given in Section~\ref{sec:data_model}. The event selection and 
categorization are described in Section~\ref{sec:event-selection}, followed by 
a discussion of the systematic uncertainties considered for the analysis of 
the data in Section~\ref{sec:uncertainties}. The results of the search are 
presented in Section~\ref{sec:results}. The paper is summarized in 
Section~\ref{sec:summary}.  

\section{The CMS detector}
\label{sec:detector}

The central feature of the CMS apparatus is a superconducting solenoid of 
$6\unit{m}$ internal diameter, providing a magnetic field of $3.8\unit{T}$. Within 
the solenoid volume are a silicon pixel and strip tracker, a lead tungstate 
crystal electromagnetic calorimeter (ECAL), and a brass and scintillator hadron 
calorimeter (HCAL), each composed of a barrel and two endcap sections. Forward 
calorimeters extend the pseudorapidity ($\eta$) coverage provided by the barrel 
and endcap detectors. Muons are detected in gas-ionization chambers embedded in 
the steel flux-return yoke outside the solenoid.

The silicon tracker measures charged particles within the range of $\abs{\eta}
<2.5$. During the LHC data-taking period up to 2017, the silicon tracker consisted 
of $1440$ silicon pixel and $15\,148$ silicon strip detector modules. From 2017 
on, the silicon pixel detector was upgraded to $1856$ modules. For nonisolated 
particles with a transverse momentum of $1 < \pt < 10\GeV$ with respect to the 
beam axis and $\abs{\eta}<1.4$, the track resolutions are typically $1.5\%$ in 
$\pt$ and $25$--$90$ ($45$--$150$)$\mum$ in the transverse (longitudinal) impact 
parameter~\cite{Chatrchyan:2014fea}. From 2017 on, the transverse impact parameter 
resolution improved to $20$--$60\mum$ when restricted to the same $\eta$ range 
as before and $20$--$75\mum$ in the increased full $\eta$ range~\cite{CMS-DP-2020-049}. 

The momentum resolution for electrons with $\pt \approx 45\GeV$ from $\ZEE$ decays 
ranges from $1.7$ to $4.5\%$. It is generally better in the barrel region than in 
the endcaps, and also depends on the bremsstrahlung energy emitted by the electron 
traversing the material in front of the ECAL~\cite{Khachatryan:2015hwa}.

Muons are measured in the range of $\abs{\eta} < 2.4$, with detection planes made 
using three technologies: drift tubes, cathode strip chambers, and resistive plate 
chambers. The relative $\pt$ resolution for muons with $20 <\pt < 100\GeV$ is 
$1.3$ to $2.0\%$ in the barrel and better than $6\%$ in the endcaps. In the barrel 
the relative $\pt$ resolution is better than $10\%$ for muons with $\pt$ up to 
$1\TeV$~\cite{Chatrchyan:2012xi}.

In the barrel section of the ECAL, an energy resolution of about $1\%$ is achieved 
for unconverted or late-converting photons in the tens of GeV energy range. The 
remaining barrel photons have a resolution of about $1.3\%$ up to $\abs{\eta}=1$, 
rising to about $2.5\%$ at $\abs{\eta} = 1.4$. In the endcaps, the resolution of 
unconverted or late-converting photons is about $2.5\%$, while the remaining endcap 
photons have a resolution of $3$--$4\%$~\cite{Khachatryan:2015iwa}. 

When combining information from the entire detector, the jet energy resolution 
amounts typically to $15$--$20\%$ at $30\GeV$, $10\%$ at $100\GeV$, and $5\%$ at 
$1\TeV$~\cite{Khachatryan:2016kdb}.

Events of interest are selected using a two-tiered trigger system. The first 
level (L1), composed of custom hardware processors, uses information from the 
calorimeters and muon detectors to select events at a rate of around 
$100\unit{kHz}$ within a fixed latency of about $4\mus$~\cite{Sirunyan:2020zal}. 
The second level, known as the high-level trigger (HLT), consists of a farm of 
processors running a version of the full event reconstruction software optimized 
for fast processing, and reduces the event rate to around $1\unit{kHz}$ before 
data storage~\cite{Khachatryan:2016bia}. 

A more detailed description of the CMS detector, together with a definition of 
the coordinate system used and the relevant kinematic variables, can be found 
in Ref.~\cite{Chatrchyan:2008zzk}. 

\section{Event reconstruction}
\label{sec:event-reconstruction}

The reconstruction of the proton-proton ($\Pp\Pp$) collision products is based on 
the particle-flow (PF) algorithm, as described in Ref.~\cite{Sirunyan:2017ulk}, 
combining the available information from all CMS subdetectors to reconstruct 
individual particle candidates, categorized into electrons, photons, muons, charged 
and neutral hadrons. The average number of interactions per bunch crossing in the 
data of the years 2016 (2017 and 2018) used in this search was $23$ ($32$). The 
fully recorded detector data of a bunch crossing defines an event for further 
processing. The candidate vertex with the largest value of summed physics-object 
$\pt^{2}$ is taken to be the primary vertex (PV) of the event. The physics objects 
for this purpose are the jets, formed using the anti-$\kt$ jet finding algorithm 
as implemented in the \FASTJET package~\cite{Cacciari:2011ma} with the 
tracks assigned to the corresponding candidate vertex as inputs, and the associated 
missing transverse momentum, taken as the negative vector sum of the $\pt$ of 
those jets. Secondary vertices, which are displaced from the PV in the transverse 
plane are indicative of decays of long lived particles emerging from the PV. Any 
other collision vertices in the event are associated with additional mostly soft 
inelastic $\Pp\Pp$ collisions called pileup (PU).

Electron candidates are reconstructed by fitting tracks in the tracker, and then 
matching the tracks to clusters in the ECAL~\cite{Khachatryan:2015hwa,Sirunyan:2020ycc}. 
To increase their purity, reconstructed electrons are required to pass a multivariate 
electron identification discriminant, which combines information on track quality, 
shower shape, and kinematic quantities. For this analysis, a working point with an 
identification efficiency of $90\%$ is used, for a rate of jets misidentified as 
electrons of ${\approx}1\%$. 

Muons in the event are reconstructed by performing a simultaneous track fit to 
hits in the tracker and in the muon detectors~\cite{Chatrchyan:2012xi,Sirunyan:2018fpa}. 
The presence of hits in the muon detectors already leads to a strong suppression 
of particles misidentified as muons. Additional identification requirements on 
the track fit quality and the compatibility of individual track segments with the 
fitted track reduce the misidentification rate further. For this analysis, muon 
identification requirements with an efficiency of ${\approx}99\%$ are chosen. 

The contributions from backgrounds to the electron and muon selections are further 
reduced by requiring the corresponding lepton to be isolated from any hadronic 
activity in the detector. This property is quantified by an isolation variable
\begin{linenomath}
  \begin{equation}
    \Irelem=\frac{1}{\ptem}\left(\sum\pt^{\text{charged}} + \max\left(0, 
    \sum\et^{\text{neutral}}+\sum\et^{\gamma}-\pt^{\text{PU}}\right)\right),
  \end{equation}
\end{linenomath}
where $\ptem$ corresponds to the electron or muon $\pt$ and $\sum\pt^{\text{charged}}$, 
$\sum\et^{\text{neutral}}$, and $\sum\et^{\gamma}$ to the $\pt$ (transverse energy 
$\et$) sum of all charged particles, neutral hadrons, and photons, in a predefined 
cone of radius $\Delta R = \sqrt{\smash[b]{\left(\Delta\eta\right)^{2}+\left(\Delta
\phi\right)^{2}}}$ around the lepton direction at the PV, where $\Delta\phi$ 
(measured in radians) and $\Delta\eta$ correspond to the angular distances of the 
particle to the lepton in the azimuthal angle $\phi$ and $\eta$ directions, 
respectively~\cite{Khachatryan:2015hwa,Chatrchyan:2012xi}. The chosen cone sizes are 
$\Delta R<0.3$ for electrons and $0.4$ for muons. The lepton itself is excluded 
from the calculation. To mitigate any distortions from PU, only those charged 
particles whose tracks are associated with the PV are taken into account. Since 
for neutral hadrons and photons an unambiguous association to the PV or PU is not 
possible, an estimate of the contribution from PU ($\pt^{\text{PU}}$) is subtracted 
from the sum of $\sum\et^{\text{neutral}}$ and $\sum\et^{\gamma}$. This estimation 
is obtained from tracks not associated to the PV in the case of $I_{\text{rel}}^{
\Pgm}$ and from the mean energy flow per unit area in the case of $I_{\text{rel}
}^{\Pe}$. In the case of negative values, the results are set to zero.

For further characterization of an event, all reconstructed PF objects are 
used to form jets using the anti-$\kt$ jet finding algorithm with a distance 
parameter of $0.4$. To identify jets resulting from the hadronization of $\PQb$ 
quarks ($\PQb$ jets) the \textsc{DeepJet} algorithm is used as described in 
Refs.~\cite{Sirunyan:2017ezt,Bols:2020bkb}. In this analysis a working point of 
this algorithm is chosen that corresponds to an expected  $\PQb$ jet identification 
efficiency of ${\approx}80\%$ for an expected  misidentification rate for jets 
originating from light quarks and gluons ($\PQc$ quarks) of $1\%\,(15\%)$~\cite{CMS-DP-2018-058}. 
Jets with $\pt>30\GeV$ and $\abs{\eta}<4.7$ and $\PQb$ jets with $\pt>20\GeV$ and 
$\abs{\eta}<2.4\,(2.5)$ are used, where the value in parentheses corresponds to 
the selection after the upgrade of the silicon pixel detector from 2017 on. 

Jets are also used as seeds for the reconstruction of hadronic $\Pgt$ decays 
($\tauh$). This is done by further exploiting the substructure of the jets, using 
the hadrons-plus-strips algorithm, as described in Ref.~\cite{Sirunyan:2018pgf}. 
For the analysis, the decays into one or three charged hadrons with up to two neutral 
pions with $\pt>2.5\GeV$ are used. The neutral pions are reconstructed as strips 
with dynamic size in $\eta$-$\phi$ from reconstructed electrons and photons 
contained in the seeding jet, where the strip size varies as a function of the 
$\pt$ of the electron or photon candidate. The $\tauh$ decay mode is then obtained 
by combining the charged hadrons with the strips. To distinguish the $\tauh$ decays 
from jets originating from the hadronization of quarks or gluons, and from electrons, 
or muons, the \textsc{DeepTau} algorithm is used, as described in Ref.~\cite{CMS-DP-2019-033}. 
This algorithm exploits the information of the reconstructed event record, comprising 
tracking, impact parameter, and ECAL and HCAL cluster information; the kinematic and 
object identification properties of the PF candidates in the vicinity of the 
$\tauh$ candidate and the $\tauh$ candidate itself; and several characterizing 
quantities of the whole event. It results in a multiclassification output $\yDT_{
\alpha}\,(\alpha=\Pgt,\,\text{jet}\,,\Pe\,,\Pgm)$ equivalent to a Bayesian 
probability of the $\tauh$ candidate to originate from a genuine tau, the 
hadronization of a quark or gluon, an isolated electron, or an isolated muon. 
From this output three discriminators are built according to 
\begin{linenomath}
  \begin{equation}
    D_{\alpha} = \frac{\yDT_{\Pgt}}{\yDT_{\Pgt}+\yDT_{\alpha}}\,, \quad
    \alpha=\text{jet},\,\Pe,\,\Pgm.
  \end{equation}
\end{linenomath}
For this analysis, a working point of $\Dj$ with a genuine $\tauh$ identification 
efficiency of $70\%$ for a misidentification rate of $0.43\%$ is chosen. For $\De$ 
and $\Dm$, depending on the $\Pgt\Pgt$ final state, different working points with 
efficiencies of $80\%$ and ${>}99\%$ and misidentification rates between 
$0.03\%$ and $2.60\%$ are chosen, respectively. It should be noted that the 
misidentification rate of $\Dj$ strongly depends on the $\pt$ and quark flavor of 
the misidentified jet, which is why this number should be viewed as an approximate 
estimate. 

The pileup-per-particle identification algorithm~\cite{Bertolini:2014bba} is 
applied to reduce the PU dependence of the $\ptvecmiss$ observable, which is 
computed as the negative vectorial $\pt$ sum of the PF candidates, weighted by 
their probability to originate from the PV~\cite{Sirunyan:2019kia}. Its magnitude 
is referred to as $\ptmiss$. It is used for the estimation of the invariant mass 
of the two tau leptons before their decay, as discussed in 
Section~\ref{sec:event-selection}.

\begin{table}[t]
  \topcaption{
    Background processes contributing to the event selection, as given in 
    Section~\ref{sec:event-selection}. The symbol $\ell$ corresponds to an 
    electron or muon. The second column refers to the experimental signature 
    in the analysis, the last three columns indicate the estimation methods 
    used to model each corresponding signature, as described in 
    Sections~\ref{sec:tau-embedding}--\ref{sec:simulation}. 
  }
  \label{tab:bg-processes}
  \centering
  \begin{tabular}{lcccc}
    & &\multicolumn{3}{c}{Estimation method} \\
    Background process & Final state signature & $\Pgt$-emb. & $\FF$ & Sim. \\
    \hline
    \multirow{3}{*}{$\PZ$} & $\Pgt\Pgt$ & $\checkmark$ & $\NA$ & $\NA$ \\
    & $\Jettau$  & $\NA$ & $\checkmark$ & $\NA$ \\
    & $\ell\ell$ & $\NA$ & $\NA$ & $\checkmark$ \\ [\cmsTabSkip]
    \multirow{3}{*}{$\ttbar$} & $\Pgt\Pgt+X$ & $\checkmark$ & $\NA$ & $\NA$ \\
    & $\Jettau$  & $\NA$ & $\checkmark$ & $\NA$ \\ 
    & $\hphantom{\Pgt}\ell+X$   & $\NA$ & $\NA$ & $\checkmark$ \\ [\cmsTabSkip]
    \multirow{3}{*}{Diboson+single $\PQt$} & $\Pgt\Pgt+X$ & $\checkmark$ & $\NA$ & $\NA$ \\
    & $\Jettau$  & $\NA$ & $\checkmark$ & $\NA$ \\
    & $\hphantom{\Pgt}\ell+X$   & $\NA$ & $\NA$ & $\checkmark$ \\ [\cmsTabSkip]
    $\PW$+jets   & $\Jettau$ & $\NA$ & $\checkmark$ & $\NA$ \\[\cmsTabSkip]
    QCD multijet & $\Jettau$ & $\NA$ & $\checkmark$ & $\NA$ \\[\cmsTabSkip]
    \multirow{2}{*}{Single $\Ph$} & $\Pgt\Pgt$ & $\NA$ & $\NA$ & $\checkmark$ \\
    & $\PQb\PQb$ & $\NA$ & $\NA$ & $\checkmark$ \\ %[\cmsTabSkip]
    \hline
    \multicolumn{5}{r}{\small{$\ell=\Pe,\,\Pgm$}} \\
  \end{tabular}
\end{table}

\section{Data model}
\label{sec:data_model}

The selection given in Section~\ref{sec:event-selection} targets the reconstruction 
of a pair of tau leptons originating from $\Ph$ with a mass of $\mtautau=125\GeV$ 
and a pair of $\PQb$ quarks originating from $\Phs$ with a mass varying between 
$60$ and $2800\GeV$. For the $\Pgt$ pair the $\etau$, $\mutau$ and $\tautau$ final 
states are used. The contribution of the $\emu$ final state to the sensitivity of 
the search has been found to be negligible, which can be understood from the low 
$\Pgt\Pgt$ branching fraction and the overwhelmingly large background from $\PQt$ 
quark pair production ($\ttbar$). In the $\etau$ and $\mutau$ final states, the 
most abundant source of background after the selection is $\ttbar$ that can easily 
result in a signature with genuine leptons and $\PQb$ quarks. After selection the 
expected fraction of $\ttbar$ events in these final states is ${\approx}70\%$. In 
the $\tautau$ final state, events containing purely quantum chromodynamics (QCD) 
induced gluon and quark jets, referred to as QCD multijet production in the following, 
and the decay of $\PZ$ bosons into tau leptons form the largest background sources 
with ${\approx}35\%$ each. 

All SM background sources of relevance for this analysis are listed in 
Table~\ref{tab:bg-processes}. For the background modeling, three different methods 
are used depending on the interpreted signature after reconstruction: (i) $\Pgt
\Pgt$ events are obtained from the $\Pgt$-embedding method, discussed in 
Section~\ref{sec:tau-embedding}; (ii) events with jets misidentified as $\tauh$ 
($\jettau$) are obtained from the $\FF$-method, discussed in Section~\ref{sec:FF-method}; 
(iii) all other background events and the signal events are obtained from full 
event simulation, discussed in Section~\ref{sec:simulation}. 

\subsection{\texorpdfstring{The $\Pgt$-embedding method}
{The tau-embedding method}}
\label{sec:tau-embedding}

For all events in which the decay of a $\PZ$ or two $\PW$ bosons results in two 
genuine tau leptons, the $\Pgt$-embedding method, as described in 
Ref.~\cite{Sirunyan:2019drn}, is used. For this purpose, $\Pgm\Pgm$ events are 
selected in data. All energy deposits of the muons are removed from the event 
record and replaced by simulated tau lepton decays with the same kinematic 
properties as the selected muons. In this way the method relies only on the 
simulation of the well-understood tau lepton decay and its energy deposits in 
the detector, while all other parts of the event, such as the reconstructed jets, 
their identification as originating from the PV, the identification of $\PQb$ 
jets, or the non-$\Pgt$ related parts of $\ptmiss$, are obtained from data. This 
obviates the need to simulate complicated processes, such as parton showering, 
hadronization, underlying event, and event pileup, which are difficult to model 
in simulation, and results in an improved description of the data compared to 
the simulation of the full process. In turn, several simulation-to-data corrections, 
as detailed in Section~\ref{sec:corrections}, are not needed. 

The selected muons predominantly originate from $\PZ$ boson decays; however, 
contributions from other processes resulting in two genuine tau leptons, like 
$\ttbar$ or diboson production, are also covered by this model, where throughout 
the text diboson refers to any combination of two $\PW$ or $\PZ$ bosons. A 
detailed discussion of the selection of the original $\Pgm\Pgm$ events, the exact 
procedure itself, its range of validity, and related uncertainties can be found 
in Ref.~\cite{Sirunyan:2019drn}. For a selection with at least one jet identified 
as a $\PQb$ jet in the event, as described in Section~\ref{sec:event-selection}, 
$84\%$ of the $\Pgm\Pgm$ events selected for the $\Pgt$-embedding method are 
expected to originate from $\PZ$ boson decays, $14\%$ from $\ttbar$, and 
${\approx}2\%$ from diboson production. 

\subsection{\texorpdfstring{The $\FF$-method}{The FF-method}}
\label{sec:FF-method}

The main contributing processes to $\jettau$ events are QCD multijet production, 
$\PW$ bosons in association with jets ($\Wjets$), and $\ttbar$. These events are 
estimated using the $\FF$-method, as described in Refs.~\cite{Sirunyan:2018qio,
Sirunyan:2018zut}. For this purpose the complete kinematic phase space is split 
into the disjoint signal region (SR), application region (AR), and determination 
regions (DR$^{i}$). The SR and the AR differ only in the working point chosen for 
the identification of the $\tauh$ candidate, where for the AR a looser working 
point is chosen and the events from the SR are excluded. Three independent 
extrapolation factors $\FF^{i}$ are derived for QCD multijet, $\Wjets$ production, 
and $\ttbar$ in three dedicated DR$^{i}$, defined to enrich each corresponding 
process. The $\FF^{i}$ are then used to estimate the yields $N_{\text{SR}}$ and 
kinematic properties of the combination of these backgrounds in the SR from the 
number of events $N_{\text{AR}}$ in the AR according to 
\begin{linenomath}
  \begin{equation}
    \label{eq:FF}
    N_{\text{SR}} = \left(\sum\limits_{i}w_{i}\FF^{i}\right)N_{\text{AR}}
    \qquad i=\text{QCD, }\PW\text{+jets, }\ttbar.
  \end{equation}
\end{linenomath}
For this purpose the $\FF^{i}$ are combined into a weighted sum, using the 
simulation-based estimation of the fractions $w_{i}$ of each process in the AR.

For the estimate of $\FF^{\text{QCD}}$, the charges of the two selected $\Pgt$ 
decay products are required to be of same sign. For the estimation of $\FF^{\Wjets}$, 
a $\PQb$ jet veto and a high transverse mass of the lepton-$\ptmiss$ system are 
required. The estimation of $\FF^{\ttbar}$ is obtained from simulation, with a 
selection of more than two jets, at least one $\PQb$ jet, and more than two leptons 
in an event. Each $\FF^{i}$ is derived on an event-by-event basis, as a function 
of the $\pt$ of the $\tauh$ candidate, the $\pt$ of the second $\Pgt$ decay in the 
event, and the mass of the visible $\Pgt\Pgt$ decay products. All other processes 
but the enriched background process are estimated from the $\tau$-embedding method 
or simulation and subtracted for this purpose. Each $\FF^{i}$ is further subject 
to a number of nonclosure corrections derived from control regions in data to 
take sub-leading dependencies of the $\FF^{i}$ into account. 

\subsection{Simulation}
\label{sec:simulation}

In the $\tautau$ final state, the $\Pgt$-embedding and $\FF$-methods cover 
${\approx}95\%$ of all expected background events. In the $\etau$ and $\mutau$ 
final states the fractions of expected background events described by these two 
methods are ${\approx}42\%$, each. All remaining events originate from processes 
like $\PZ$ boson, $\ttbar$, or diboson production, where at least one decay of 
a vector boson into an electron or muon is not covered by any of the two methods. 
These and the signal events are modeled using the simulation of the full 
processes. 

The production of $\PZ$ bosons in the $\Pe\Pe$ and $\Pgm\Pgm$ final states is
simulated at leading-order (LO) precision in the coupling strength 
$\alpS$, using the \MGvATNLO 2.2.2 (2.4.2) event generator~\cite{Alwall:2011uj,
Alwall:2014hca} for the simulation of the data taken in 2016 (2017 and 2018). 
To increase the number of simulated events in regions of high signal purity, 
supplementary samples are generated with up to four outgoing partons in the hard 
interaction. For diboson production \MGvATNLO is used at next-to-LO (NLO) 
precision in $\alpS$. For $\ttbar$ and single~$\PQt$ quark production samples 
are generated at NLO precision using \POWHEG 2.0~\cite{Nason:2004rx,
Frixione:2007vw,Alioli:2008tz,Alioli:2010xd,Alioli:2010xa,Bagnaschi:2011tu}. The 
kinematic properties of single $\Ph$ production are simulated at NLO precision 
using \POWHEG separately for the production via gluon fusion, vector boson 
fusion, or in association with a $\PZ$ boson, $\PW$ boson, or a top quark pair. 
For this purpose $\Ph$ is assumed to behave as expected from the SM.

When compared to data, $\PZ$ boson, $\ttbar$, and single~$\PQt$ quark events in 
the $\PQt\PW$ channel are normalized to their cross sections at next-to-NLO 
precision in $\alpS$~\cite{Melnikov:2006kv,Czakon:2011xx,Kidonakis:2013zqa}. 
Single~$\PQt$ quark production in the $t$-channel and diboson events are normalized 
to their cross sections at NLO precision in $\alpS$ or higher~\cite{Kidonakis:2013zqa,
Campbell:2011bn,Gehrmann:2014fva}.

The signal process $\PH\to\Ph\Phs$ is generated using \MGvATNLO at LO precision. 
The analysis is restricted to $\PH$ production via gluon fusion, which is expected 
to be dominant, \eg, in the NMSSM. Due to the two unknown masses involved in the 
decay, a two-dimensional grid of signal mass pairs is generated, resulting in 
$420$ mass pairs spanning from $240$ to $3000\GeV$ in $\mH$ and $60$ to $2800
\GeV$ in $\mhs$, only taking pairs with $\mhs+125\GeV\leq\mH$ into account.

{\tolerance=800
For the generation of all signal and background processes, the 
NNPDF3.0~\cite{Ball:2014uwa} (NNPDF3.1~\cite{Ball:2017nwa}) parton distribution 
functions are used for the simulation of the data taken in 2016 (2017 and 2018). 
The description of the underlying event is parameterized according to the 
CUETP8M1~\cite{Khachatryan:2015pea} and CP5~\cite{Sirunyan:2019dfx} tunes. 
Parton showering and hadronization, as well as the $\Pgt$ lepton decays, are 
modeled using the \PYTHIA 8.230 event generator~\cite{Sjostrand:2014zea}. For 
all simulated events, additional inclusive inelastic $\Pp\Pp$ collisions generated 
with \PYTHIA are added according to the expected PU profile in data to take the 
effect of the observed PU into account. All events generated are passed through 
a \GEANTfour-based~\cite{Agostinelli:2002hh} simulation of the CMS detector and 
reconstructed using the same version of the CMS event reconstruction software as 
used for the data.
\par}

\subsection{Corrections and control of the model}
\label{sec:corrections}

The capability of the model to describe the data is monitored in various control 
regions orthogonal to the signal and background classes defined in 
Section~\ref{sec:event-selection}, and corrections and corresponding uncertainties 
are derived where necessary. 

The following corrections equally apply to simulated and $\Pgt$-embedded events, 
where the $\Pgt$ decay is also simulated. Since the simulation part for $\Pgt
$-embedded events happens under detector conditions, which are different from 
the case of fully simulated events, corrections and related uncertainties may 
differ, as detailed in Ref.~\cite{Sirunyan:2019drn}. Corrections are derived for 
residual differences between data and simulation in the efficiency of the selected 
triggers, the electron and muon tracking efficiency, and in the efficiency of the 
identification and isolation requirements for electrons and muons. These corrections 
are obtained in bins of $\pt$ and $\eta$ of the corresponding lepton, using the 
``tag-and-probe'' method, as described in Ref.~\cite{Khachatryan:2010xn}, with 
$\ZEE$ and $\ZMM$ events. They usually amount to not more than a few percent. 

In a similar way, corrections are obtained for the efficiency of triggering on the 
$\tauh$ decay signature and for the $\tauh$ identification efficiency, following 
procedures as described in Ref.~\cite{Sirunyan:2018pgf}. The latter are derived 
as a function of the $\pt$ of the $\tauh$ in four bins below $40\GeV$ and one bin 
above. For $\pt(\tauh)>40\GeV$, a correction is also derived for each $\tauh$ 
decay mode individually, which is used only in the $\tautau$ final state. 
Corrections to the energy scale of the $\tauh$ decays and of electrons misidentified 
as $\tauh$ are derived for each year of data-taking and each $\tauh$ decay mode 
individually, from likelihood scans of discriminating observables, like the mass 
of the visible decay products of the $\tauh$ candidate, as detailed in 
Ref.~\cite{Sirunyan:2018pgf}. For muons misidentified as $\tauh$ this effect has 
been observed to be negligible. For the trigger efficiency the correction is 
obtained from parametric fits to the trigger efficiency as a function of $\pt$ 
derived for each corresponding sample and data.

The following corrections only apply to fully simulated events. During the 2016 
and 2017 data-taking, a gradual shift in the timing of the inputs of the ECAL L1 
trigger in the region at $\abs{\eta} > 2.0$ caused a specific trigger inefficiency. 
For events containing an electron (a jet) with $\pt$ larger than ${\approx}50\,
({\approx}100)\GeV$, in the region of $2.5<\abs{\eta}<3.0$ the efficiency loss is 
$10$--$20\%$, depending on $\pt$, $\eta$, and time. Corresponding corrections have 
been derived from data and applied to the simulation. 

The jet energy is corrected to the expected response at the stable hadron level,
using corrections measured in bins of the jet $\pt$ and 
$\eta$, as described in Ref.~\cite{Khachatryan:2016kdb}. These corrections are 
usually not larger than $10$--$15\%$. Residual data-to-simulation corrections are 
applied to the simulated event samples. They usually range between subpercent 
level at high jet $\pt$ in the central part of the detector to a few percent in 
the forward region. A correction is applied to the direction and magnitude of 
$\ptvecmiss$ based on differences between estimates of the hadronic recoil in 
$\ZMM$ events in data and simulation, as described in Ref.~\cite{Sirunyan:2019kia}. 
This correction is applied to the simulated $\PZ$ boson, single $\Ph$, and signal 
events, where a hadronic recoil against a single particle is well defined. 

The efficiencies for genuine and misidentified $\PQb$ jets to pass the working 
points of the $\PQb$ jet identification discriminator, as given in 
Section~\ref{sec:event-reconstruction}, are determined from data, using $\ttbar$ 
events for genuine $\PQb$ jets and $\PZ$ boson production in association with jets 
originating from light quarks or gluons. Data-to-simulation corrections are obtained 
for these efficiencies and used to correct the number of $\PQb$ jets in the 
simulation, as described in Ref.~\cite{Sirunyan:2017ezt}. 

Data-to-simulation corrections are further applied to $\ZEE$ ($\ZMM$) events in 
the $\etau$ ($\mutau$) and $\tautau$ final states in which an electron (muon) 
is reconstructed as a $\tauh$ candidate, to account for residual differences in 
the $\Pe(\Pgm)\to\tauh$ misidentification rate between data and simulation. 
Deficiencies in the modeling of $\PZ$ boson events in the $\Pe\Pe$, $\Pgm\Pgm$ 
final states, due to the use of a LO simulation, are corrected for by reweighting 
the simulated $\ZMM$ events to data in bins of $\pt^{\Pgm\Pgm}$ and $m_{\Pgm\Pgm}$. 
In addition all simulated $\ttbar$ events are weighted to better match the top 
quark $\pt$ distribution, as observed in data~\cite{Khachatryan:2015oqa}. 

The overall normalization of all backgrounds is constrained by dedicated event 
categories, obtained from neural network (NN) multiclassification, as described 
in Section~\ref{sec:event-selection}. After the event selection and prior to the 
event classification, \ie, still at an inclusive state of the analysis, the 
marginal distributions and pairwise correlations, including self-correlations, of 
all input features to the NNs used for event classification are subject to 
extensive scrutiny. This is done exploiting goodness-of-fit tests, based on a 
saturated likelihood model~\cite{Baker:1983tu} including all systematic uncertainties 
of the model and their correlations, as used for the signal extraction. This 
guarantees a good understanding of the input space to the NNs and the input 
distributions used for the statistical inference of the signal contribution. 

\section{Event selection and classification}
\label{sec:event-selection}

\subsection{Event selection}

The L1 trigger decision is based on the identification of high-$\pt$ electrons 
or muons, reconstructed from a fast readout of the ECAL and muon detectors. A 
positive L1 trigger decision initiates the further reconstruction of the given 
event at the HLT. In the HLT step, the selection is based on the presence of a 
single electron or muon, an $\etau$ or $\mutau$ pair, or a $\tauh\tauh$ pair in 
the event. The addition of the single-electron or single-muon requirement to the 
list of triggers via a logical OR condition increases the overall acceptance of 
the online selection. In the offline selection further requirements on the $\pt$, 
$\eta$, $\Irelem$, and the $D_{\alpha}$ discriminators are applied in addition 
to the object identification requirements described in 
Section~\ref{sec:event-reconstruction}, as summarized in 
Table~\ref{tab:selection_kin}.

\begin{table}[b]
  \centering
  \topcaption
  {
    Offline requirements applied to electrons, muons, and $\tauh$ candidates used 
    for the selection of the $\Pgt$ pair. The $\pt$ values in parentheses correspond 
    to events selected by a single-electron or single-muon trigger. These requirements 
    depend on the year of data-taking. For $\Dj$ the efficiency and for $D_{\Pe(
    \Pgm)}$ the misidentification rates for the chosen working points are given 
    in parentheses. A detailed discussion is given in the text. 
  }
  \label{tab:selection_kin}
  \begin{tabular}{lll}
    Final state & Electron/Muon &  $\tauh$ \\ 
    \hline
    $\etau$ & $\pt>25\,(26,\,28,\,33)\GeV$ & $\pt>35\,(30)\GeV$ \\ 
    & $\abs{\eta}<2.1$ & $\abs{\eta}<2.3$ \\ 
    &	$I_{\text{rel}}^{\Pe}<0.15$ & $\Dj\,(70\%)$, $\De\,(0.05\%)$, $\Dm\,(0.13\%)$ \\ [\cmsTabSkip]
    $\mutau$ & $\pt>20\,(23,\,25)\GeV$ & $\pt>35\,(30)\GeV$ \\
    & $\abs{\eta}<2.1$ & $\abs{\eta}<2.3$ \\ 
    & $I_{\text{rel}}^{\Pgm}<0.15$ & $\Dj\,(70\%)$, $\De\,(2.60\%)$, $\Dm\,(0.03\%)$ \\[\cmsTabSkip]
    $\tautau$ & $\hspace{0.7cm}\NA$ & $\pt>40\GeV$ \\
    & & $\abs{\eta}<2.1$ \\ 
    & & $\Dj\,(70\%)$, $\De\,(2.60\%)$, $\Dm\,(0.13\%)$ \\ 
    \hline
  \end{tabular}
\end{table}

In the $\etau$ ($\mutau$) final state, an electron (muon) with at least $25\,(20)
\GeV$ is required, if an event was selected by a trigger based on the presence 
of the $\etau$ ($\mutau$) pair. If the event was selected by a single-electron 
trigger, the $\pt$ requirement on the electron is increased to $26$--$33\GeV$ 
depending on the data-taking period, to ensure a sufficiently high efficiency of 
the HLT selection. For muons, the $\pt$ requirement is increased to $23\,(25)
\GeV$ for 2016 (2017 or 2018), if selected by a single-muon trigger. The electron 
(muon) is required to be contained in the central detector with $\abs{\eta}<2.1$, 
and to be isolated from any hadronic activity according to $\Irelem<0.15$. The 
$\tauh$ candidate is required to have $\abs{\eta}<2.3$ and $\pt>35\GeV$ if 
selected by an $\etau\,(\mutau)$ pair trigger, or $\pt>30\GeV$ if selected by a 
single-electron (single-muon) trigger. In the $\tautau$ final state, both $\tauh$ 
candidates are required to have $\abs{\eta}<2.1$ and $\pt>40\GeV$. The working 
points of the \textsc{DeepTau} discriminants, as described in 
Section~\ref{sec:event-reconstruction}, are chosen depending on the final state. 
Events with additional leptons fulfilling looser selection criteria are discarded 
to avoid the assignment of single events to more than one final state.

The selected $\Pgt$ decay candidates are required to be of opposite charge and 
to be separated by more than $\Delta R = 0.5$ in the $\eta$-$\phi$ plane. The 
closest distance of their tracks to the PV is required to be $d_{z}<0.2\cm$ 
along the beam axis. For electrons and muons, an additional requirement of $d_{xy}
<0.045\cm$ in the transverse plane is applied. In rare cases in which an extra 
$\tauh$ candidate fulfilling all selection requirements is found, the candidate 
with the higher score of $\Dj$ is chosen.

In addition to the tau lepton pair, at least one $\PQb$ jet fulfilling the 
selection criteria, as described in Section~\ref{sec:event-reconstruction}, is 
required. Events that contain only one $\PQb$ jet and no other jet are removed 
from the analysis. If more than two $\PQb$ jets exist, the pair is built from 
those that are leading in $\pt$. If only one $\PQb$ jet exists the $\PQb$ pair 
is built using the $\PQb$ jet and the jet with its highest $\PQb$ jet score of 
the \textsc{DeepJet} classifier. The energies of the jets used for the $\PQb$ 
pair are corrected using the multivariate energy-momentum regression described 
in Ref.~\cite{Sirunyan:2019wwa}. 

This analysis selection is optimized for the reconstruction of events where the 
$\Ph$ and $\Phs$ decay products are spatially resolved. Boosted topologies, which 
can occur in parts of the explored mass ranges, are not specifically targeted.

\subsection{Event classification}

All events retained by the selection described above are further sorted into five 
categories. One for signal, the other four are enriched with different backgrounds. 
This is done separately for each of the three final states and each of the three 
data-taking periods resulting in 45 categories. The background-enriched categories 
are used to further constrain systematic uncertainties in the background estimates 
during the statistical inference of the signal contribution. This categorization 
is based on NN multiclassification exploiting fully connected feed-forward NNs 
with two hidden layers of $200$ nodes each, and five output nodes implemented in 
the software package \textsc{TensorFlow}~\cite{tensorflow2015-whitepaper}. The 
first four output nodes used to enrich the backgrounds comprise the following 
events: (i) events containing genuine $\Pgt$ pairs (labeled ``$\Pgt\Pgt$''); (ii) 
events with quark or gluon induced jets misidentified as $\tauh$ (labeled 
``$\jettau$''); (iii) top quark pair events where the intermediate $\PW$ bosons 
decay into any combination of electrons and muons, or into a single $\Pgt$ and 
an electron or muon (not included in (i) or (ii); labeled as ``\ttc''); (iv) 
events from remaining background processes that are of minor importance for the 
analysis and not yet included in any of the previous classes (labeled as ``misc''). 
The processes in (iv) comprise diboson production, single~$\PQt$ quark production, 
$\PZ$ boson decays to electrons or muons, and single $\Ph$ production. For single 
$\Ph$ production, rates and branching fractions as predicted by the SM are assumed. 
Event classes (i) and (ii) are defined by final state or experimental signature 
of the contained events rather than explicit underlying physics processes. They 
are complemented by event classes (iii) and (iv) to characterize all background 
processes, which are of relevance for the analysis. The fifth event class, associated 
with the fith output node, contains the $\PH\to\Ph(\Pgt\Pgt)\Phs(\PQb\PQb)$ signal 
events (labeled as ``signal''). This choice of event classes closely resembles the 
data model described in Section~\ref{sec:data_model}. 

For each node in the hidden layers, the hyperbolic tangent is chosen as the 
activation function. The activation function for the output layer is chosen to 
be the softmax function allowing for a Bayesian conditional probability 
interpretation $y_{i}^{(k)}$ of an event $k$ to be associated to an event class 
$i$, given its input features $\vec{x}^{(k)}$ to the NN. The highest value of 
$y_{i}^{(k)}$, $\max(y_{i}^{(k)})$, defines which class the event is associated 
with and will define the discriminator for the statistical inference of the signal 
contribution. All other outputs $y_{j}^{(k)},\,j\neq i$ are discarded from any 
further consideration so that any event is used only once for the statistical 
inference of the signal. 

In the $\etau$ and $\mutau$ final states, the input space to the NNs is spanned by 
$20$ features $\vec{x}$ of an event including $\pt$ of the $\tau$ candidates and 
the jets forming the $\PQb$ quark pair; the mass and $\pt$ estimates of the $\Pgt$ 
pair, $\PQb$ quark pair, and $\Pgt\Pgt\PQb\PQb$ system; the number of ($\PQb$) 
jets; and further kinematic properties of the selected jets. For this purpose, a 
likelihood-based estimate of the $\Pgt\Pgt$ mass before decay~\cite{Bianchini:2014vza} 
and a kinematic fit to the $\Pgt\Pgt\PQb\PQb$ system for each given $\mH$ and 
$\mhs$ hypothesis, similar to the approach described in Ref.~\cite{Khachatryan:2015tha}, 
are used. In the $\tautau$ final state these features are complemented by the 
masses of the two jets used for the $\PQb$ quark pair system and their associated 
output values of the \textsc{DeepJet} algorithm, to allow for a better discrimination 
of genuine $\PQb$ jets from light quark or gluon induced jets. All input features 
have been selected from a superset of variables describing the properties of the 
event exploiting a ranking of individual features and pairwise correlations of 
features, as described in Ref.~\cite{Wunsch:2018oxb}. 

Since the kinematic properties of the signal strongly vary across the probed 
ranges of $\mH$ and $\mhs$ a total of $68$ NNs per final state are used 
for classification, which within each final state only differ by the kinematic 
properties of the signal that are used for training. For this purpose, adjacent 
sets of points in $\mhs$ and $\mH$ are combined into single signal classes. Up 
to four points in $\mhs$ are combined for single points in $\mH$, for $\mH\leq
1000\GeV$. Beyond $\mH=1000\GeV$, all remaining points in $\mH$ and up to nine 
points in $\mhs$ are combined. The concrete grouping is a tradeoff between 
sensitivity and computational feasibility. Though it reduces the use of the 
invariant mass of the reconstructed $\PQb$ quark pair ($\mbb$) for the NN decision 
this grouping of mass points has only a small effect on the overall NN performance 
in separating signal from background, which can be understood by the following 
means: (i) correlated information, like the $\mH$ estimate and the $\chi^{2}$ of 
the kinematic fit are used, in addition to $\mbb$; (ii) the fact that $\mbb$ is 
a peaking distribution for signal while not for background is still fully exploited 
by the NN; (iii) for $\mH>1000\GeV$ the $\pt$ of the jets forming the $\PQb$ quark 
pair gains importance. Differences of the input features depending on the year 
of data-taking are taken into account by a conditional training using a one-hot 
encoding of the data-taking year in the NN training, such that the correct year 
of data taking obtains the value $1$, while all other data-taking years obtain 
the value $0$.   

The parameters to be optimized during training are the weights ($\{w_{a}\}$) and 
biases ($\{b_{b}\}$) of the NN output functions $y_{i}$. Before training the 
weights are initialized with random numbers using the Glorot initialization 
technique~\cite{glorot2010} with values drawn from a uniform distribution. The 
biases are initialized with zero. The trainings are then performed using randomly 
sampled batches of $N=30$ events per event class, drawn from the training datasets 
using a balanced batch approach~\cite{Shimizu2018}. This approach has shown 
improved convergence properties on training samples with highly imbalanced 
lengths. The classification task is encoded in the NN loss function, chosen as 
the cross entropy
\begin{linenomath}
  \begin{equation}
    \begin{split}
      &L\left(\{y_{i}^{(k)}\},\{y^{\prime(k)}_{j}\}\right) = 
      -\sum\limits_{k=1}^{N}
      \,y^{\prime(k)}_{j}
      \log\left(y_{i}^{(k)}(\{w_{a}\},\{b_{b}\},\{\vec{x}^{(k)}\})\right)\,;
      \qquad y_{j}^{\prime(k)}=\delta_{ij}, \\
    \end{split}
    \label{eq:loss-function}
  \end{equation}
\end{linenomath}
which is to be minimized during the NN trainings. In Eq.~(\ref{eq:loss-function}), 
$k$ runs over the events in the batch, on which $L$ is evaluated. The NN prediction 
for event $k$ to belong to category $i$ is given by $y_{i}^{(k)}$. The function 
$y_{j}^{\prime(k)}$ encodes the prior knowledge of the training. It is $1$ if 
class $i$ of event $k$ coincides with the true event class $j$, and $0$ otherwise. 
The $y_{i}^{(k)}$ depend on the weights, biases, and input features $\{\vec{x}^{(k)}\}$ 
of event $k$ to the NN. The batch definition guarantees that each true event class 
enters the training with equal weight in the evaluation of $L$, \ie, without 
prevalence. Within the misc event class all contained processes are normalized 
according to their expected rates with respect to each other. On each batch a 
gradient step is applied, defined by the partial derivatives of $L$ in each 
weight, $w_{a}$, and bias, $b_{b}$, using the Adam minimization 
algorithm~\cite{Kingma:2014vow}, with a constant multiplicative learning rate of 
$10^{-4}$. To guarantee statistical independence, those events that are used for 
training are not used for any other step of the analysis. 

The performance of the NNs during training is monitored by evaluating $L$ on a 
validation subset that contains a fraction of $25\%$ of randomly chosen events 
from the training sample, which are excluded from the gradient computation. The 
training is stopped if the evaluation of $L$ on the validation dataset does not 
indicate any further decrease for a sequence of $50$ epochs, where an epoch is 
defined by $1000$ ($100$) batches in the $\etau$/$\mutau$ ($\tautau$) final 
state. The NNs used for the analysis are then defined by the weights and biases of 
the epoch with the minimal value of $L$ on the validation sample. 

To improve the generalization property of the NNs, two regularization techniques 
are introduced. Firstly, after each hidden layer a layer with a dropout probability 
of $30\%$ is added. Secondly, the weights of the NNs are subject to an L2 (Tikhonov) 
regularization~\cite{Tikhonov:1963} with a regularization factor of $10^{-5}$.

After training, a very good separation between the background events and the 
signal events is achieved, with a purity and classification sensitivity for the 
correct signal class of typically more than $80\%$.

\section{Systematic uncertainties}
\label{sec:uncertainties}

The uncertainty model used for the analysis comprises theoretical uncertainties,  
experimental uncertainties, and uncertainties due to the limited population of 
template distributions for the background model used for the statistical inference 
of the signal, as described in Section~\ref{sec:results}. The last group of 
uncertainties is incorporated for each bin of each corresponding template 
individually following the approach proposed in Ref.~\cite{Barlow:1993dm}. For 
this analysis, where the signal is expected to be concentrated to a few bins 
with low background expectation, these uncertainties can often range among those 
with the largest impact on the signal significance. All other uncertainties lead 
to correlated changes across bins either in the form of normalization changes or 
as general nontrivial shape-altering variations. Depending on the way they are 
derived, correlations may also arise across years, samples, or individual 
uncertainties.  

The following uncertainties related to the level of control of the reconstruction 
of electrons, muons, and $\tauh$ candidates after selection apply to simulated 
and $\Pgt$-embedded events. Unless stated otherwise they correspond to the 
uncertainties of the corrections described in Section~\ref{sec:corrections} and 
are partially correlated across $\Pgt$-embedded and simulated events. 

\begin{itemize}
\item 
  Uncertainties in the identification efficiency of electrons and muons amount to 
  $2\%$, correlated across all years. Since no significant dependence on the $\pt$ 
  or $\eta$ of each corresponding lepton is observed these uncertainties are 
  introduced as normalization uncertainties. 

\item
  With a similar reasoning, uncertainties in the electron and muon trigger 
  efficiencies are also introduced as normalization uncertainties. They amount 
  to $2\%$ each. Due to differences in the trigger leg definitions they are 
  treated as uncorrelated for single-lepton and two-object triggers. This may 
  result in shape-altering effects in the overall model, since both triggers act 
  on different regions in lepton $\pt$.
  
\item 
  For fully simulated events an uncertainty in the electron energy scale is derived 
  from the calibration of ECAL crystals, and applied on an event-by-event 
  basis~\cite{Khachatryan:2015hwa}. For $\Pgt$-embedded events uncertainties of 
  $0.50$--$1.25\%$, split by the ECAL barrel and endcap regions, are derived for 
  the corrections described in Section~\ref{sec:corrections}. Due to the different 
  ways the uncertainties are determined and differences in detector conditions 
  they are treated as uncorrelated across simulated and $\Pgt$-embedded events. 
  They lead to shape-altering variations and are treated as correlated across 
  years. The muon momentum ($p_{\Pgm}$) is very precisely known~\cite{Sirunyan:2018fpa}. 
  A variation within the given uncertainties, depending on the muon $\eta$ and 
  $\pt$ has been checked to have no influence on the analysis. 

\item 
  Uncertainties in the $\tauh$-identification range between $3$ and $5\%$ in bins 
  of $\tauh$ $\pt$. Due to the nature of how they are derived these uncertainties 
  are statistically dominated and therefore treated as uncorrelated across decay 
  modes, $\pt$ bins, and years. The same is true for the uncertainties in the 
  $\tauh$-energy scale, which range from $0.2$ to $1.1\%$, depending on the 
  $\pt$ and the decay mode of the $\tauh$. For the energy scale of electrons 
  misidentified as $\tauh$ candidates, extra corrections are derived depending 
  on the $\tauh$ $\pt$ and decay mode. Their uncertainties range from $1.0$ to 
  $2.5\%$. Concerning correlations the same statements apply as for the $\tauh
  $-energy scale. All uncertainties discussed here for the $\tauh$ identification 
  and energy scale lead to shape-altering variations. A generous variation of the 
  momentum scale of muons misidentified as $\tauh$ has been checked to have a 
  marginal effect on the analysis.

\item 
  Uncertainties in the $\tauh$ trigger efficiency are $5$--$10\%$, depending on 
  the $\pt$ of the $\tauh$. They are obtained from parametric fits to data and 
  simulation, and lead to shape-altering effects. They are treated as uncorrelated 
  across triggers and years. 
\end{itemize}

Two further sources of uncertainty are considered for $\Pgt$-embedded 
events~\cite{Sirunyan:2019drn}: 

\begin{itemize}
\item 
  A $4\%$ normalization uncertainty accounts for the level of control in the 
  efficiency of the $\Pgm\Pgm$ selection in data, which is unfolded during the 
  $\Pgt$-embedding procedure. The dominant part of this uncertainty originates 
  from the trigger used for selection and is treated as uncorrelated across 
  years. 

\item
  Another shape and normalization-altering uncertainty in the yield of $\ttbar\to
  \Pgm\Pgm+\text{X}$ decays, which are part of the $\Pgt$-embedded event samples, 
  ranges between subpercent and $10\%$, depending on the event composition of the 
  model. For this uncertainty, the number and shape of $\ttbar$ events contained 
  in the $\Pgt$-embedded event samples are estimated from simulation, for which 
  the corresponding decay has been selected at the parton level. This estimate is 
  then varied by $\pm10\%$. 
\end{itemize}

For fully simulated events the following additional uncertainties apply: 

\begin{itemize}
\item 
  Uncertainties in the $\Pe(\Pgm)\to\tauh$ misidentification rate amount to $40\%$ 
  for electrons and range from $10$ to $70\%$ for muons. The relatively large 
  size of these uncertainties originates from the rareness of these cases in the 
  control regions that are used to measure these rates. They only apply to 
  simulated $\ZEE\,(\Pgm\Pgm)$ events, which are of marginal importance for the 
  analysis. The impact on the overall background yield is below the $1\%$ level 
  both in the $\etau$ and $\mutau$ final states. The same is true for the 
  uncertainty in the reweighting in the $\PZ$ boson mass and $\pt$, discussed in 
  Section~\ref{sec:corrections}, which ranges from $10$ to $20\%$. 

\item 
  Uncertainties in the energy calibration and resolution of jets are applied with 
  different correlations depending on their sources, comprising statistical 
  limitations of the measurements used for calibration, the time-dependence of 
  the energy measurements in data due to detector aging, and nonclosure 
  corrections introduced to cover residual differences between simulation and 
  data~\cite{Khachatryan:2016kdb}. They range between subpercent level and 
  $\mathcal{O}(10\%)$, depending on the kinematic properties of the jets in the 
  event. Similar uncertainties are applied for the identification rates for $\PQb$ 
  jets and for the misidentification rates for light quark or gluon induced jets, 
  which are of a similar range each~\cite{Sirunyan:2017ezt,Bols:2020bkb}. 

\item 
  Depending on the process in consideration, two independent uncertainties in 
  $\ptmiss$ are applied. For processes that are subject to recoil corrections, 
  \ie, $\PZ$ boson production, $\Ph$ production, or signal, uncertainties 
  in the calibration and resolution of the hadronic recoil are applied, ranging 
  from $1$ to $5\%$. For all other processes an uncertainty in $\ptmiss$ is 
  derived from the amount of unclustered energy in the event~\cite{Sirunyan:2019kia}.

\item 
  A normalization uncertainty due to the timing shift of the inputs of the ECAL 
  L1 trigger described in Section~\ref{sec:corrections} amounts to $2$--$3\%$.

\item 
  A shape-altering uncertainty is derived in the reweighting of the top quark 
  $\pt$ described in Section~\ref{sec:corrections} by applying the correction 
  with twice the required magnitude, thus overcorrecting, or not applying it 
  at all. This uncertainty has only a very small effect on the final discriminator.

\item 
  The integrated luminosity is measured for each year of data-taking individually 
  following procedures, as described in Ref.~\cite{CMS:2021xjt}. The luminosity 
  measurements are known to a precision of $2.3$ $(2.5)\%$ for 
  2017~\cite{CMS-PAS-LUM-17-004} (2016~\cite{CMS-PAS-LUM-17-001} and 
  2018~\cite{CMS-PAS-LUM-18-002}). The corresponding normalization uncertainties 
  comprise parts that are correlated and parts that are uncorrelated across the 
  years.

\item 
  Uncertainties in the predictions of the normalizations of all simulated 
  processes amount to $6\%$ for $\ttbar$~\cite{Czakon:2011xx,Kidonakis:2013zqa}, 
  $5\%$ for diboson and single~$\PQt$ production~\cite{Kidonakis:2013zqa,
  Campbell:2011bn,Gehrmann:2014fva}, $2\%$ for $\PZ$ boson 
  production~\cite{Melnikov:2006kv}, and $1.3$--$3.9\%$ for the SM Higgs boson 
  production rates used for $\Ph$ production, depending on the production 
  mechanism~\cite{Alioli:2008tz,Bagnaschi:2011tu,Nason:2009ai,Luisoni:2013cuh,
  Hartanto:2015uka}. All these uncertainties are correlated across years.

\item 
  Since the search is not conducted within any particular model, no uncertainties 
  on the production cross section or branching fractions of the signal need to 
  be taken into account. Uncertainties in the signal acceptance are obtained from 
  variations of the factorization and renormalization scales, as well as from 
  sampling all relevant parameters for the estimation of the parton density 
  distributions within their corresponding uncertainties, following procedures 
  as outlined in~\cite{deFlorian:2016spz}. The changes in acceptance due to the 
  scale variations are observed to be less than $10\%$. They are shape altering, 
  depending on the $\Ph$ and $\Phs$ $\pt$. The acceptance variations due to the 
  sampling of the parton density distributions amount to normalization changes 
  of $18\%$. Both uncertainties are correlated across years.
\end{itemize}

For the $\FF$-method the following uncertainties apply: 

\begin{itemize}
\item
  The $\FF^{i}$ and their corrections are subject to statistical fluctuations in 
  each corresponding DR$^{i}$. The corresponding uncertainties are split into a 
  normalization and a shape-altering part and propagated into the final 
  discriminator. They usually range between $3$--$5\%$ and are treated as 
  uncorrelated across the kinematic and topological bins they are derived in.

\item 
  Additional uncertainties are applied to cover corrections for non-closure 
  effects and extrapolation factors, varying from a few percent to $\mathcal{O}
  (10\%)$, depending on the kinematic properties of the $\tauh$ candidate and 
  the topology of the event. These are both normalization and shape-altering 
  uncertainties.

\item
  An additional source of uncertainty concerns the subtraction of processes other 
  than the enriched process in each corresponding DR$^{i}$. These are subtracted 
  from the data using simulated or $\Pgt$-embedded events. The combined shape of 
  the events to be removed is varied by $7\%$, and the measurements are repeated. 
  The impacts of these variations are then propagated to the final discriminator 
  as shape-altering uncertainties. 

\item
  An uncertainty in the estimation of the three main background fractions in the 
  AR is estimated from a variation of each individual contribution by $7\%$, 
  increasing or decreasing the remaining fractions such that the sum of all 
  contributions remains unchanged. The amount of variation is motivated by the 
  uncertainty in the production cross sections and acceptances of the involved 
  processes and the constraint on the process composition that can be clearly 
  obtained from the AR. The effect of this variation is observed to be very small, 
  since usually one of the contributions dominates the event composition in the 
  AR. 
\end{itemize}

Due to their mostly statistical nature and differences across years, all uncertainties 
related to the $\FF$-method are treated as uncorrelated across years. A summary of 
all systematic uncertainties that have been discussed in this section is given in 
Table~\ref{tab:uncertainties}. 

\begin{table}[t]
  \topcaption{
    Summary of systematic uncertainties discussed in the text. The first column 
    indicates the source of uncertainty; the second the processes that it applies 
    to; the third the variation; and the last how it is correlated with other 
    uncertainties. A checkmark is given also for partial correlations. More 
    details are given in the text.
  }
  \label{tab:uncertainties}
  \centering
  \begin{tabular}{llcccr@{$-$}lcc}
    & & \multicolumn{3}{c}{Process} & \multicolumn{2}{c}{ } & \multicolumn{2}{c}{Correlated across} \\
    \multicolumn{2}{l}{Uncertainty} & Sim. & $\Pgt$-emb. & $\FF$ & \multicolumn{2}{c}{Variation} 
    & Years & Processes \\
    \hline 
    \multirow{2}{*}{$\Pgt$-emb.} & Acceptance & $\NA$ & $\checkmark$ & $\NA$ 
    & \multicolumn{2}{c}{$4\%$} & $\NA$ & $\NA$ \\
    & $\ttbar$ fraction & $\NA$ & $\checkmark$ & $\NA$ 
    & $0.1$ & $10\%$ & $\NA$ & $\NA$ \\[\cmsTabSkip]
    \multirow{3}{*}{$\Pgm$} 
    & Id      & $\checkmark$ & $\checkmark$ & $\NA$ 
    & \multicolumn{2}{c}{$2\%$} & $\checkmark$ & $\checkmark$ \\
    & Trigger & $\checkmark$ & $\checkmark$ & $\NA$ 
    & \multicolumn{2}{c}{$2.0\%$} & $\NA$ & $\checkmark$ \\
    & $p_{\Pgm}$ scale & $\checkmark$ & $\checkmark$ & $\NA$ 
    &  $0.1$ & $2.0\%$ & $\checkmark$ & $\checkmark$ \\[\cmsTabSkip]
    \multirow{3}{*}{$\Pe$} 
    & Id      & $\checkmark$ & $\checkmark$ & $\NA$ 
    & \multicolumn{2}{c}{$2\%$} & $\checkmark$ & $\checkmark$ \\
    & Trigger & $\checkmark$ & $\checkmark$ & $\NA$ 
    & \multicolumn{2}{c}{$2\%$} & $\NA$ & $\checkmark$ \\
    & $E_{\Pe}$ scale & $\checkmark$ & $\checkmark$ & $\NA$ 
    & \multicolumn{2}{c}{See text} & $\checkmark$ & $\checkmark$ \\[\cmsTabSkip]
    \multirow{3}{*}{$\tauh$} 
    & Id      & $\checkmark$ & $\checkmark$ & $\NA$ 
    & $3$ & $5\%$ & $\NA$ & $\checkmark$ \\
    & Trigger & $\checkmark$ & $\checkmark$ & $\NA$ 
    & $5$ & $10\%$ & $\NA$ & $\checkmark$ \\
    & $E_{\tauh}$ scale & $\checkmark$ & $\checkmark$ & $\NA$ 
    & $0.2$ & $1.1\%$ & $\NA$ & $\checkmark$ \\[\cmsTabSkip]
    \multirow{2}{*}{$\Pgm\to\tauh$} & Miss-Id & $\checkmark$ & $\NA$ & $\NA$ 
    & $10$ & $70\%$ & $\NA$ & $\NA$ \\
    & $E_{\tauh}$ scale & $\checkmark$ & $\NA$ & $\NA$ 
    & \multicolumn{2}{c}{$2\%$} & $\NA$ & $\NA$ \\[\cmsTabSkip]
    \multirow{2}{*}{$\Pe\to\tauh$} & Miss-Id & $\checkmark$ & $\NA$ & $\NA$ 
    & \multicolumn{2}{c}{$40\%$} & $\NA$ & $\NA$ \\
    & $E_{\tauh}$ scale & $\checkmark$ & $\NA$ & $\NA$ 
    & $1.0$ & $2.5\%$ & $\NA$ & $\NA$ \\[\cmsTabSkip]
    \multicolumn{2}{l}{$\PZ$ boson $\pt$ reweighting} & $\checkmark$ & $\NA$ & $\NA$
    & $10$ & $20\%$ & $\checkmark$ & $\NA$ \\[\cmsTabSkip]
    \multicolumn{2}{l}{$E_{\text{Jet}}$ scale \& resolution} & $\checkmark$ & $\NA$ & $\NA$
    & $0.1$ & $10\%$ & $\checkmark$ & $\checkmark$ \\[\cmsTabSkip]
    \multicolumn{2}{l}{$\PQb$-jet (miss-)Id} & $\checkmark$ & $\NA$ & $\NA$
    & $1$ & $10\%$ & $\NA$ & $\checkmark$ \\[\cmsTabSkip]
    \multicolumn{2}{l}{$\ptmiss$ calibration} & $\checkmark$ & $\NA$ & $\NA$
    & \multicolumn{2}{c}{See text} & $\checkmark$ & $\checkmark$ \\[\cmsTabSkip]
    \multicolumn{2}{l}{ECAL timing shift} & $\checkmark$ & $\NA$ & $\NA$
    & $2$ & $3\%$ & $\checkmark$ & $\checkmark$ \\[\cmsTabSkip]
    \multicolumn{2}{l}{$\PQt$ quark $\pt$ reweighting} & $\checkmark$ & $\NA$ & $\NA$
    & \multicolumn{2}{c}{See text} & $\checkmark$ & $\NA$ \\[\cmsTabSkip]
    \multicolumn{2}{l}{Luminosity} & $\checkmark$ & $\NA$ & $\NA$
    & $2.3$ & $2.5\%$ & $\checkmark$ & $\checkmark$ \\[\cmsTabSkip]
    \multicolumn{2}{l}{Process normalizations} & $\checkmark$ & $\NA$ & $\NA$
    & \multicolumn{2}{c}{See text} & $\checkmark$ & $\NA$ \\[\cmsTabSkip]
    \multicolumn{2}{l}{Signal acceptance} & $\checkmark$ & $\NA$ & $\NA$
    & $18$ & $20\%$ & $\checkmark$ & $\NA$ \\[\cmsTabSkip]
    \multirow{4}{*}{$\FF$} & Statistics & $\NA$ & $\NA$ & $\checkmark$
    & $3$ & $5\%$ & $\NA$ & $\NA$ \\
    & Non-closure & $\NA$ & $\NA$ & $\checkmark$
    & \multicolumn{2}{c}{$10\%$} & $\NA$ & $\NA$ \\
    & Non-$\FF$ processes & $\NA$ & $\NA$ & $\checkmark$
    & \multicolumn{2}{c}{$7\%$} & $\NA$ & $\NA$ \\
    & $\FF$ proc. composition & $\NA$ & $\NA$ & $\checkmark$
    & \multicolumn{2}{c}{$7\%$} & $\NA$ & $\NA$ \\
    \hline 
    \end{tabular}
\end{table}

\section{Results}
\label{sec:results}

The model used to infer the signal from the data is defined by an extended binned 
likelihood of the form
\begin{linenomath}
  \begin{equation}
    \mathcal{L}
    =\prod\limits_{i}\mathcal{P}(k_{i}|\mu\,S_{i}
    (\mH,\,\mhs,\,\{\theta_{j}\})+B_{i}(\{\theta_{j}\}))
    \,
    \prod\limits_{j}\mathcal{C}(\hat{\theta}_{j}|\theta_{j}),
    \label{eq:likelihood}
  \end{equation}
\end{linenomath}
where $i$ labels all bins of the distributions of the NN output functions $\max(
y_{i})$ of each of the five signal and background classes defined in 
Section~\ref{sec:event-selection}. Split by three $\Pgt\Pgt$ final states and 
three years of data-taking this results in $45$ individual input histograms, for 
each given pair of $\mH$ and $\mhs$. The function $\mathcal{P}(k_{i}|\mu\,S_{i}(
\mH,\,\mhs,\,\{\theta_{j}\})+B_{i}(\{\theta_{j}\}))$ corresponds to the Poisson 
density to observe $k_{i}$ events in bin $i$ for a prediction of $S_{i}$ signal 
and a total of $B_{i}$ background events. The parameter $\mu$ is a single scaling 
parameter of the signal. 

Systematic uncertainties are incorporated as penalty terms for additional nuisance 
parameters $\{\theta_{j}\}$ in the likelihood, appearing as a product with 
predefined probability density functions $\mathcal{C}(\hat{\theta}_{j}|\theta_{j
})$ to obtain a maximum likelihood estimate $\hat{\theta}_{j}$ for an assumed true 
value of $\theta_{j}$, during the statistical inference of the 
signal~\cite{CMS-NOTE-2011-005}.

\begin{figure}[h!]
  \centering
  \includegraphics[width=0.45\textwidth]{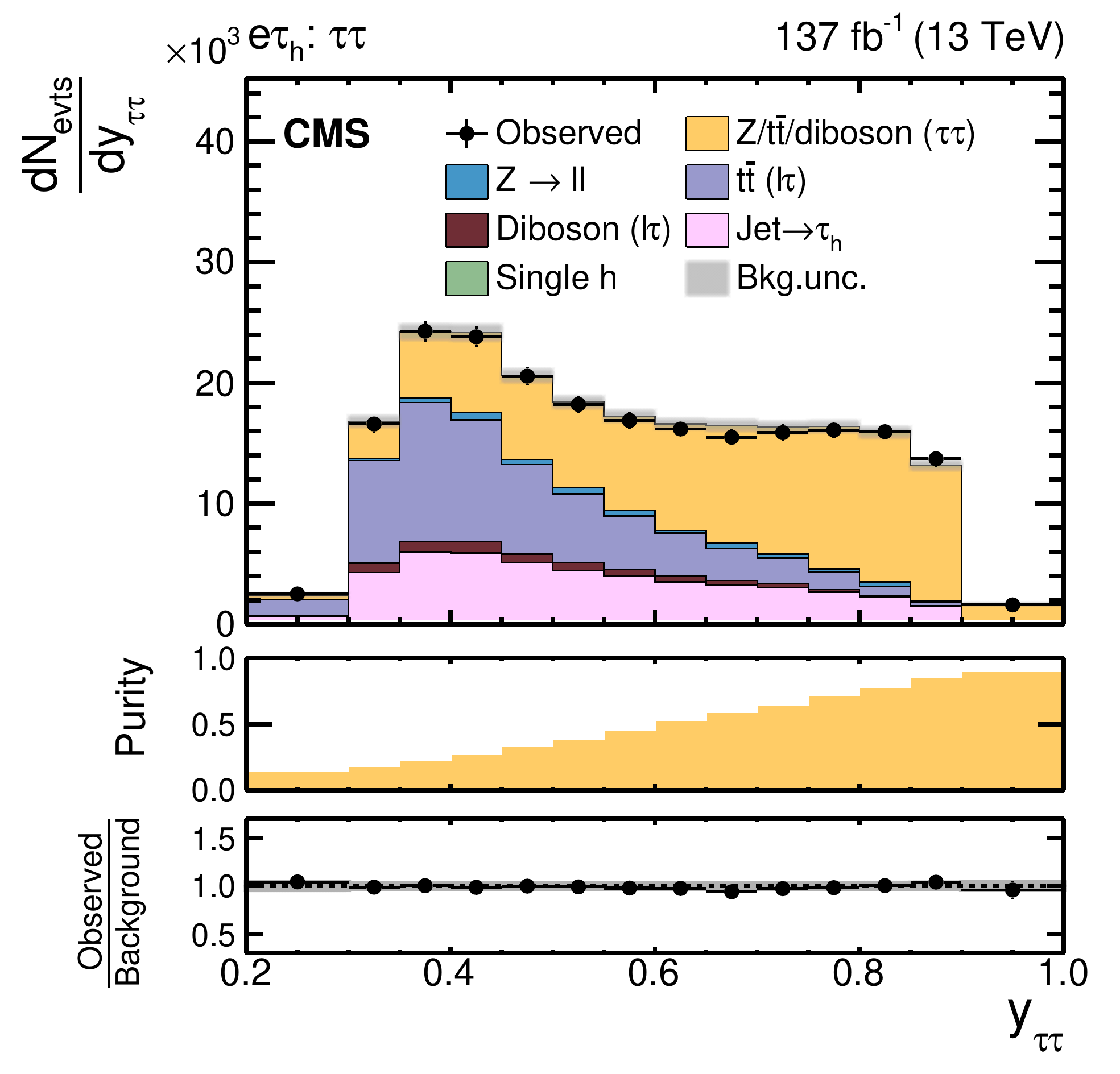}
  \includegraphics[width=0.45\textwidth]{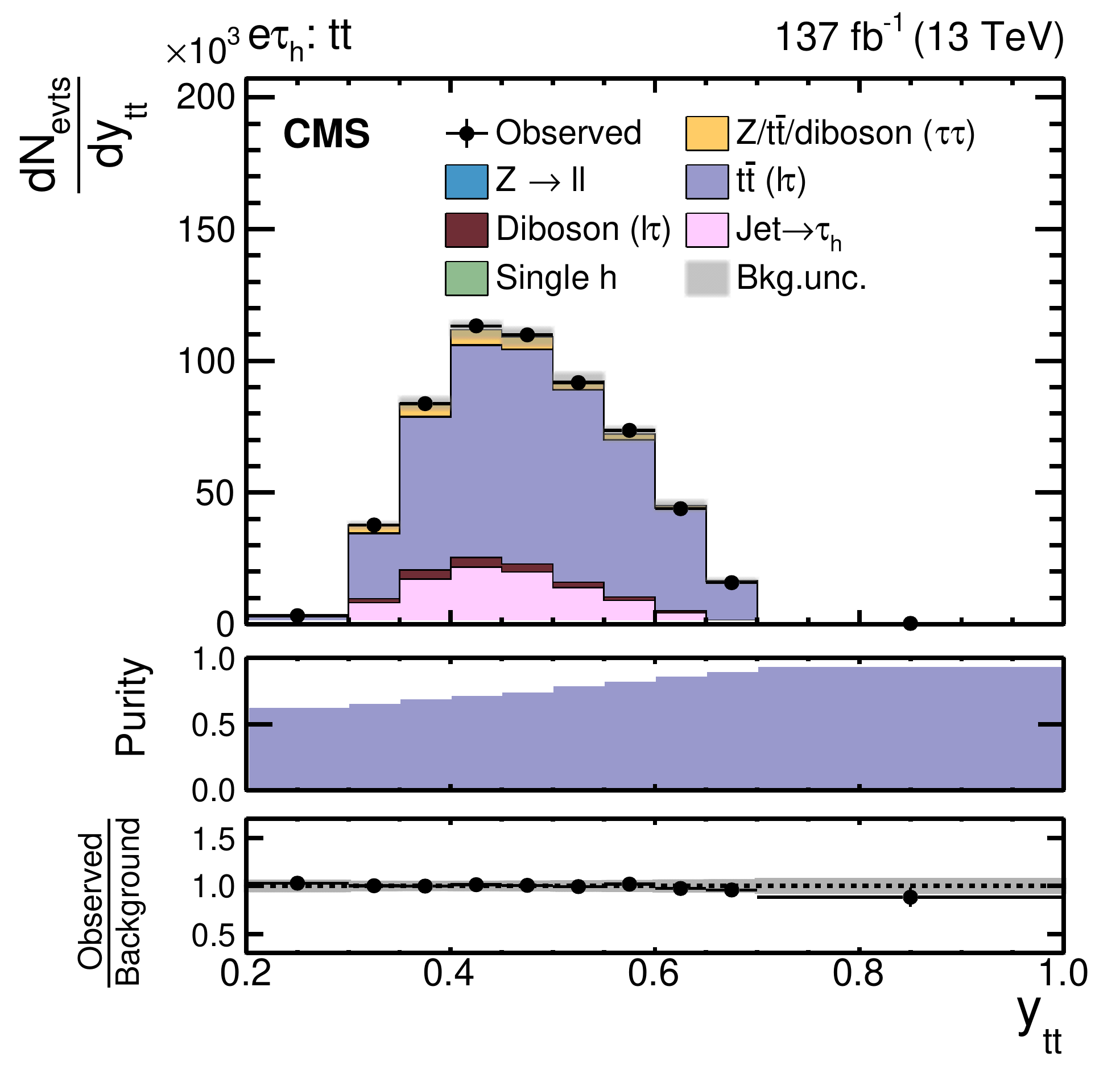}
  \includegraphics[width=0.45\textwidth]{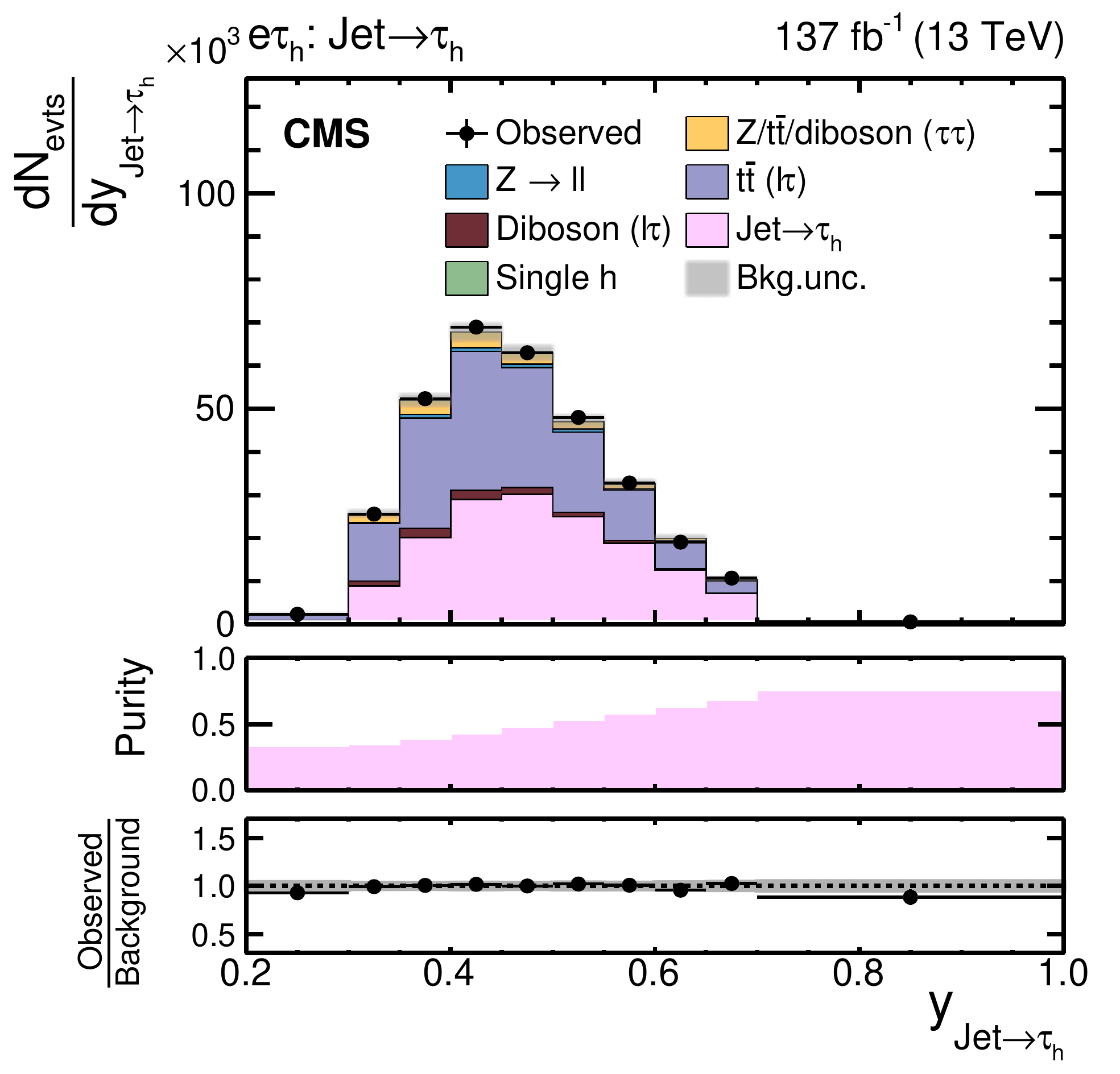}
  \includegraphics[width=0.45\textwidth]{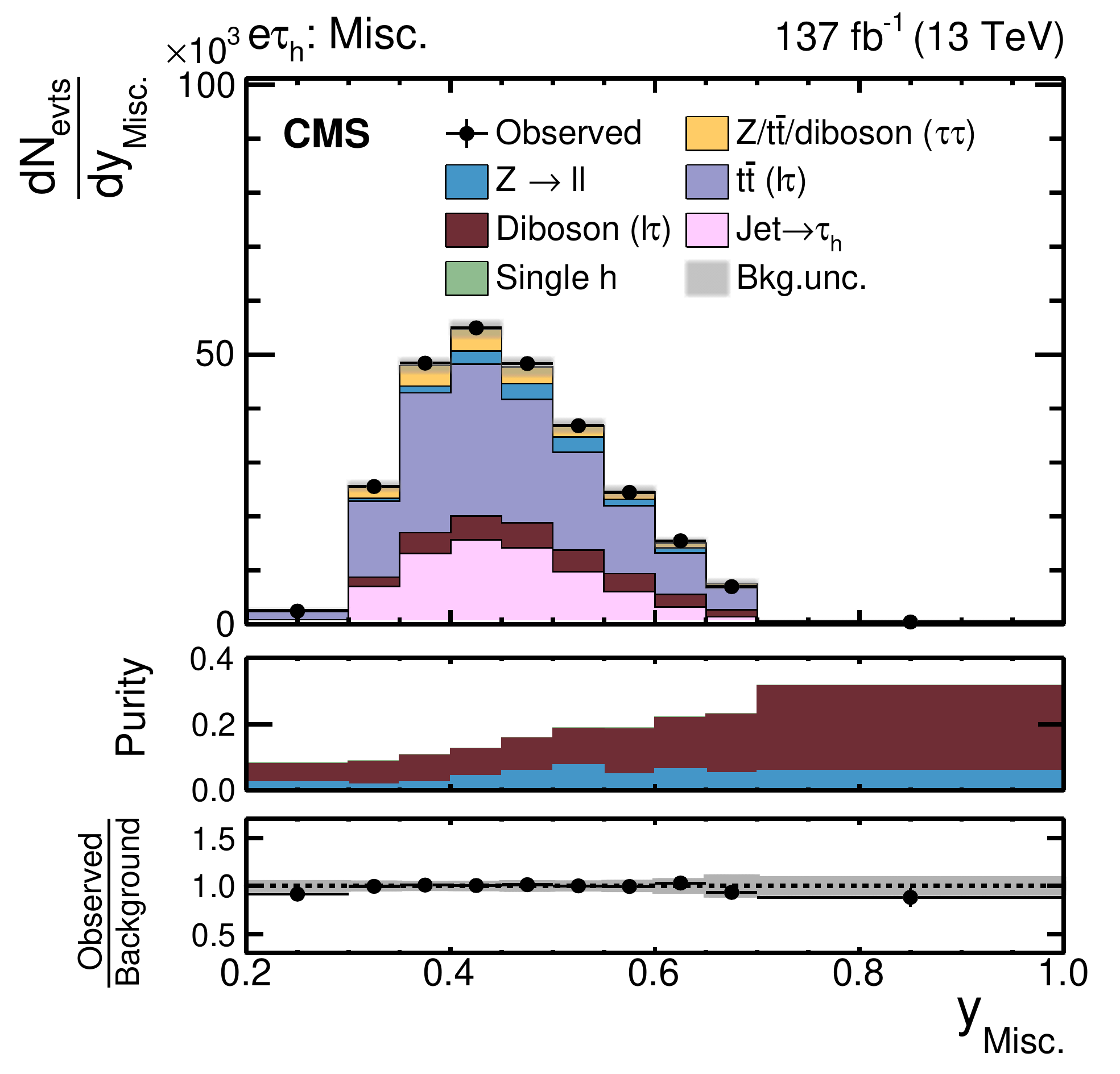}
  \includegraphics[width=0.45\textwidth]{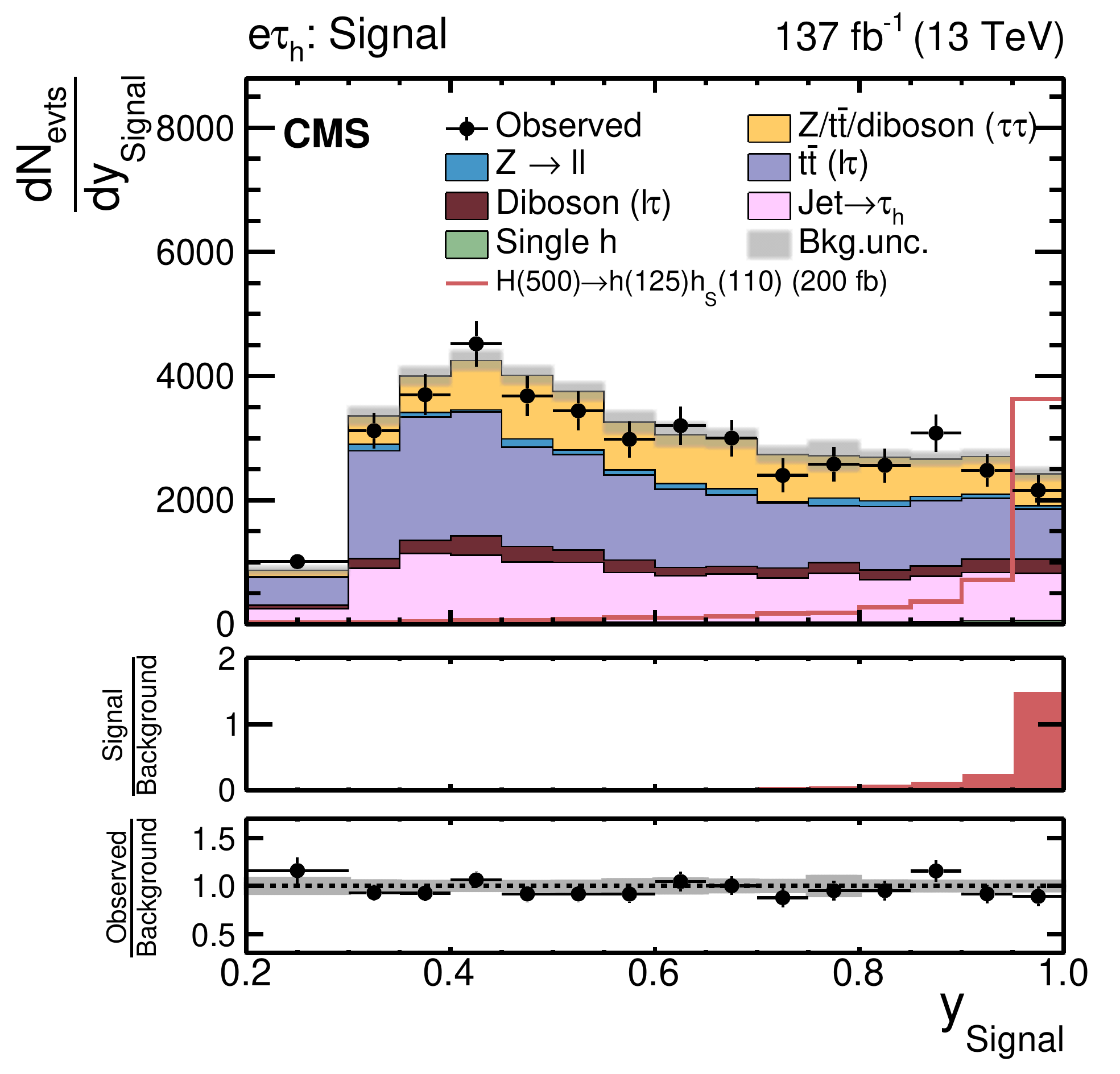}
  \hspace*{0.45\textwidth}
  \caption{
    Event categories after NN classification based on a training for $\mH=
    500\GeV$ and $100\leq \mhs<150\GeV$ in the $\etau$ final state. Shown 
    are the (upper left) $\Pgt\Pgt$, (upper right) \ttc, (middle left) $\jettau$, 
    (middle right) misc, and (lower left) signal categories. For these figures 
    the data sets of all years have been combined. The uncertainty bands correspond 
    to the combination of statistical and systematic uncertainties after the fit 
    to the signal plus background hypothesis for $\mH=500\GeV$ and $\mhs
    =110\GeV$. 
  }
  \label{fig:postfit_mt}
\end{figure}

\begin{figure}[h!]
  \centering
  \includegraphics[width=0.45\textwidth]{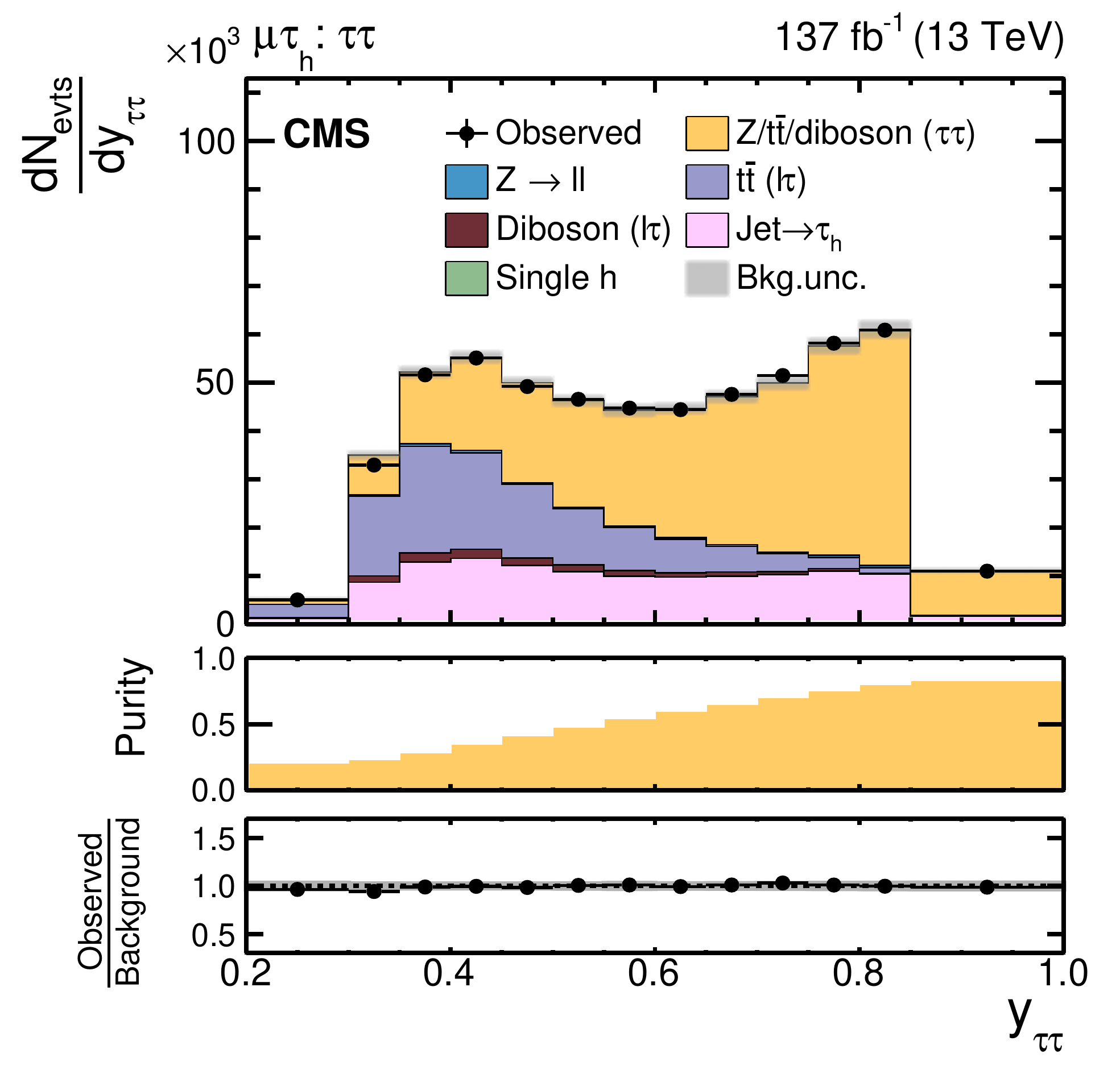}
  \includegraphics[width=0.45\textwidth]{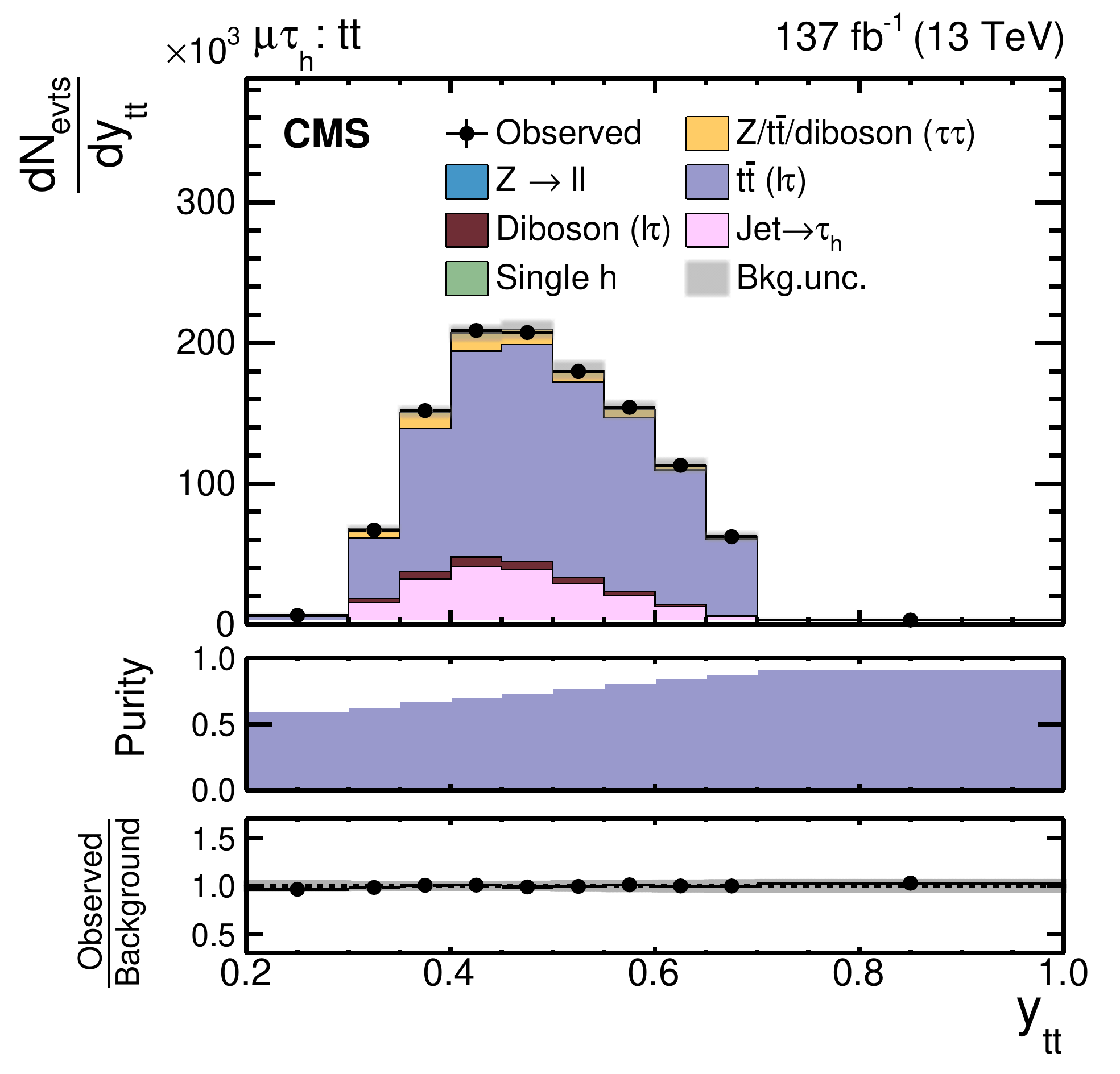}
  \includegraphics[width=0.45\textwidth]{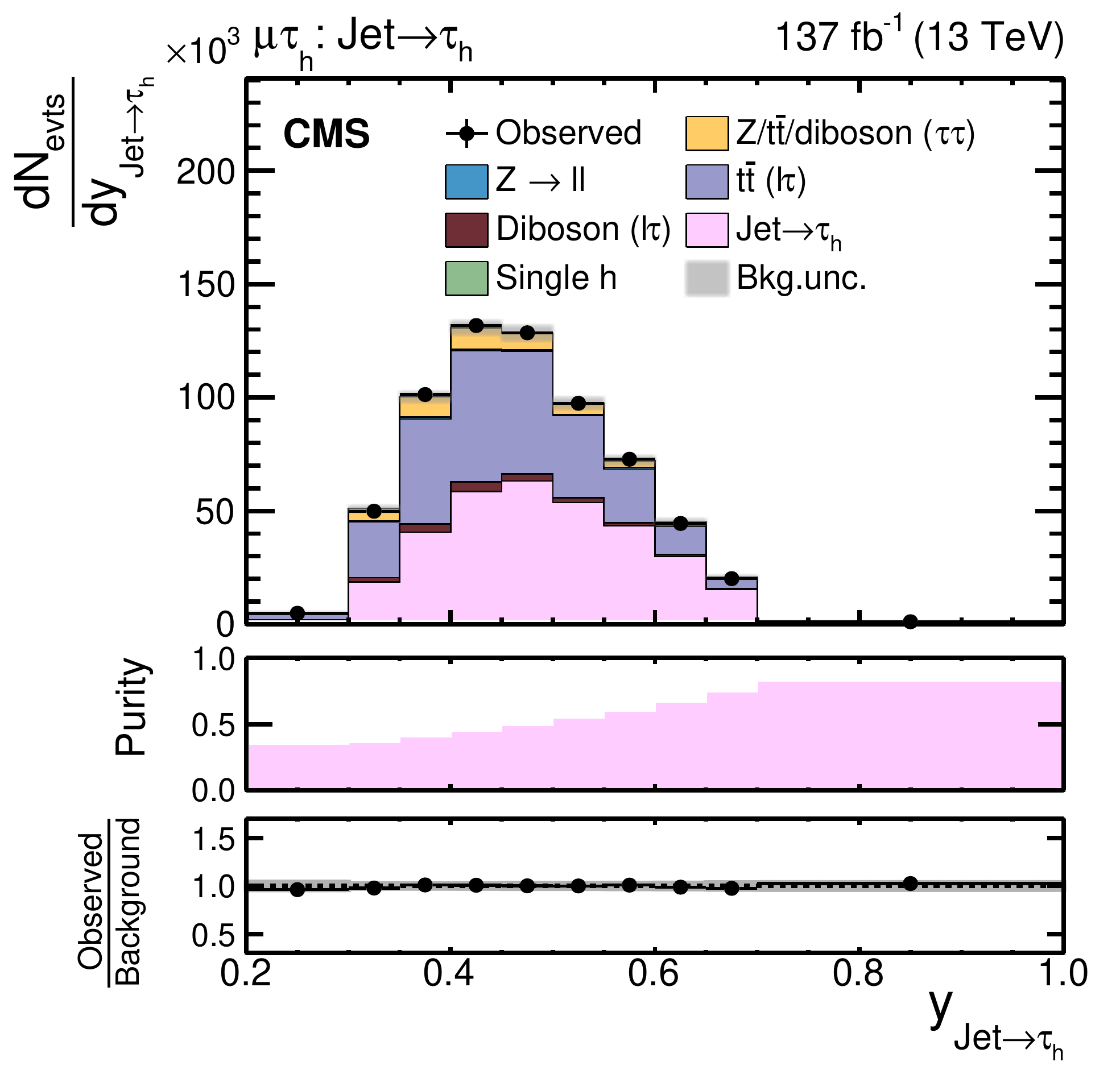}
  \includegraphics[width=0.45\textwidth]{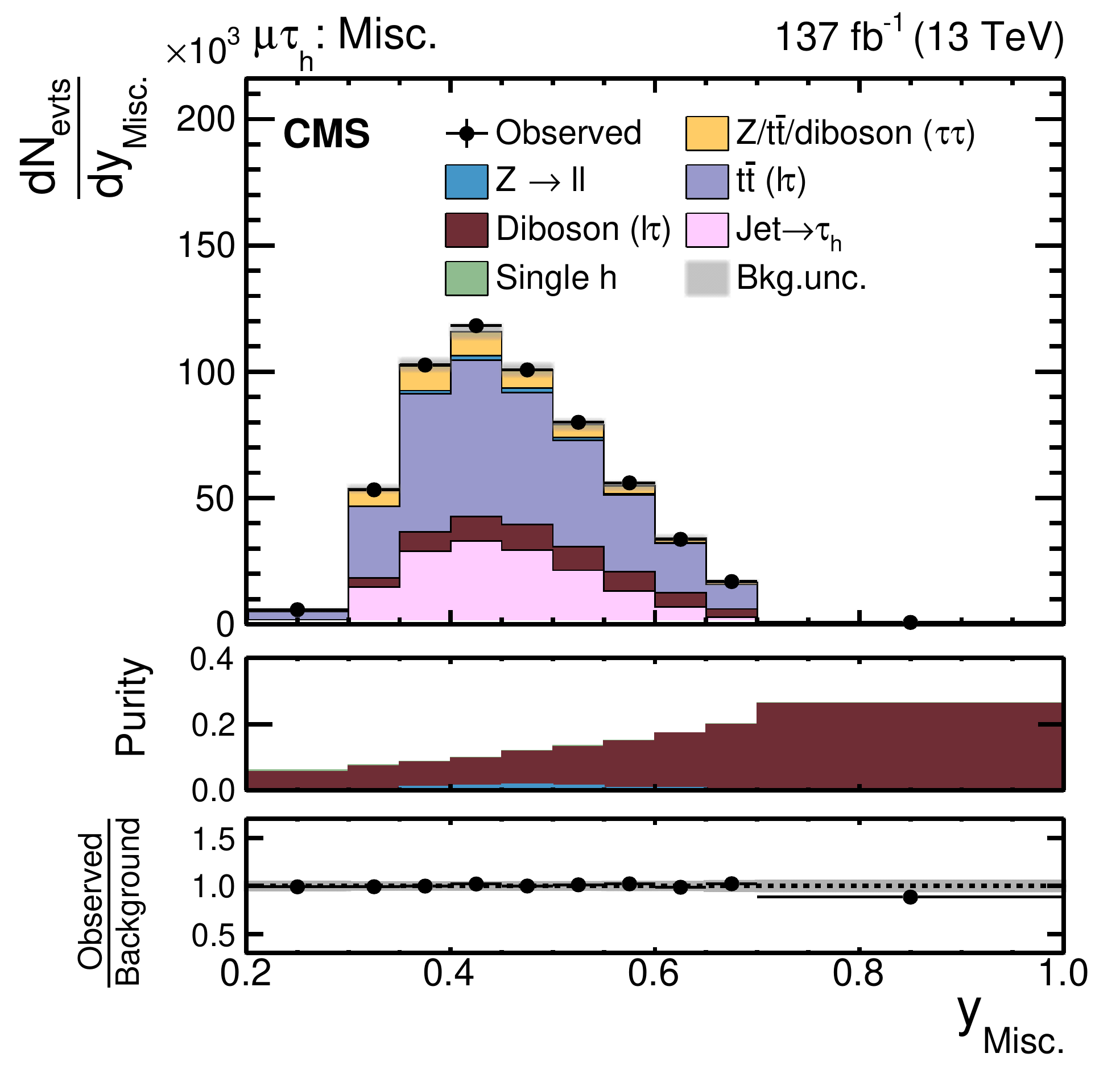}
  \includegraphics[width=0.45\textwidth]{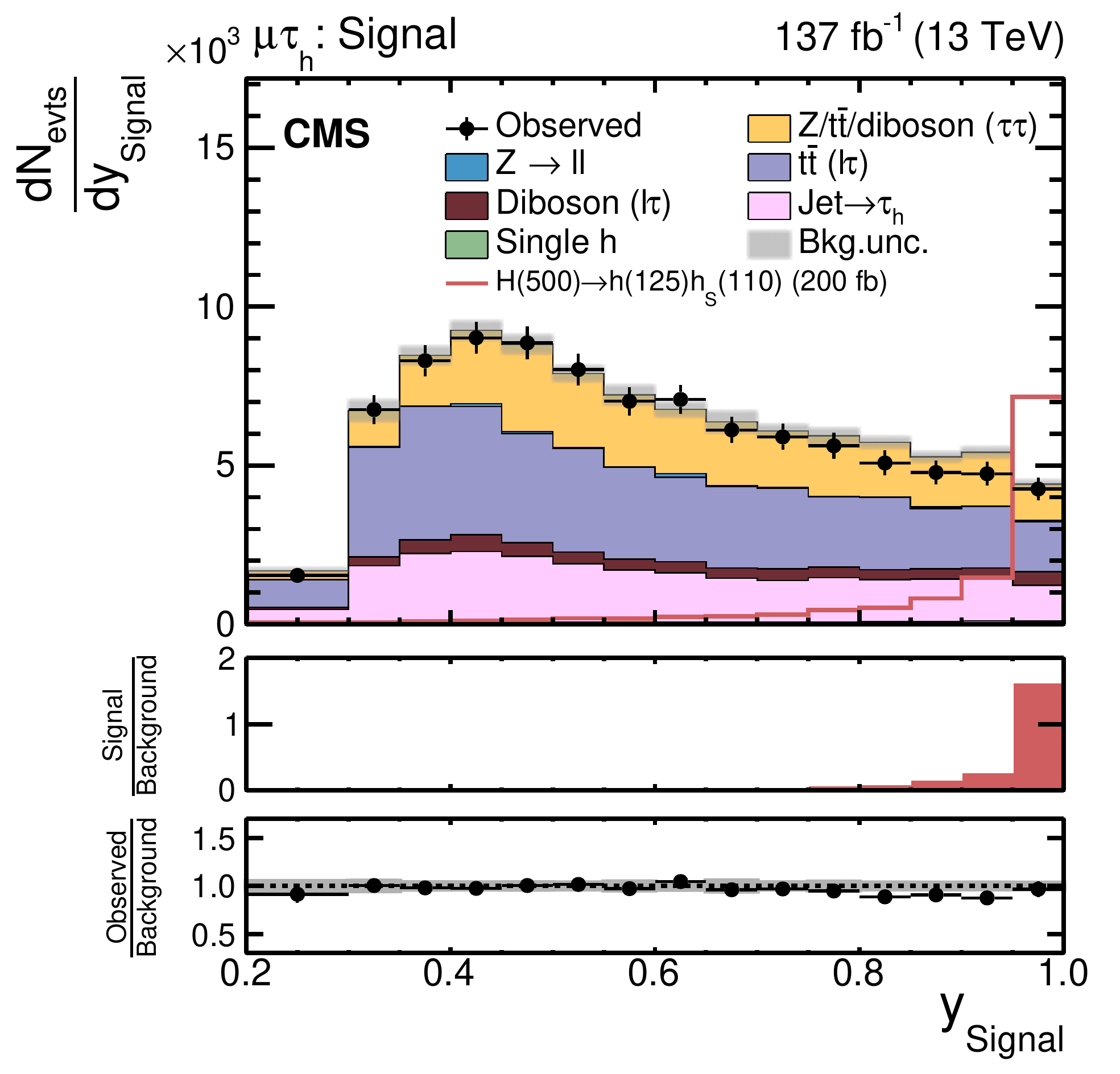}
  \hspace*{0.45\textwidth}
  \caption{
    Event categories after NN classification based on a training for $\mH=
    500\GeV$ and $100\leq \mhs<150\GeV$ in the $\mutau$ final state. Shown 
    are the (upper left) $\Pgt\Pgt$, (upper right) \ttc, (middle left) $\jettau$, 
    (middle right) misc, and (lower left) signal categories. For these figures 
    the data sets of all years have been combined. The uncertainty bands correspond 
    to the combination of statistical and systematic uncertainties after the fit 
    to the signal plus background hypothesis for $\mH=500\GeV$ and $\mhs
    =110\GeV$. 
  }
  \label{fig:postfit_et}
\end{figure}

\begin{figure}[h!]
  \centering
  \includegraphics[width=0.45\textwidth]{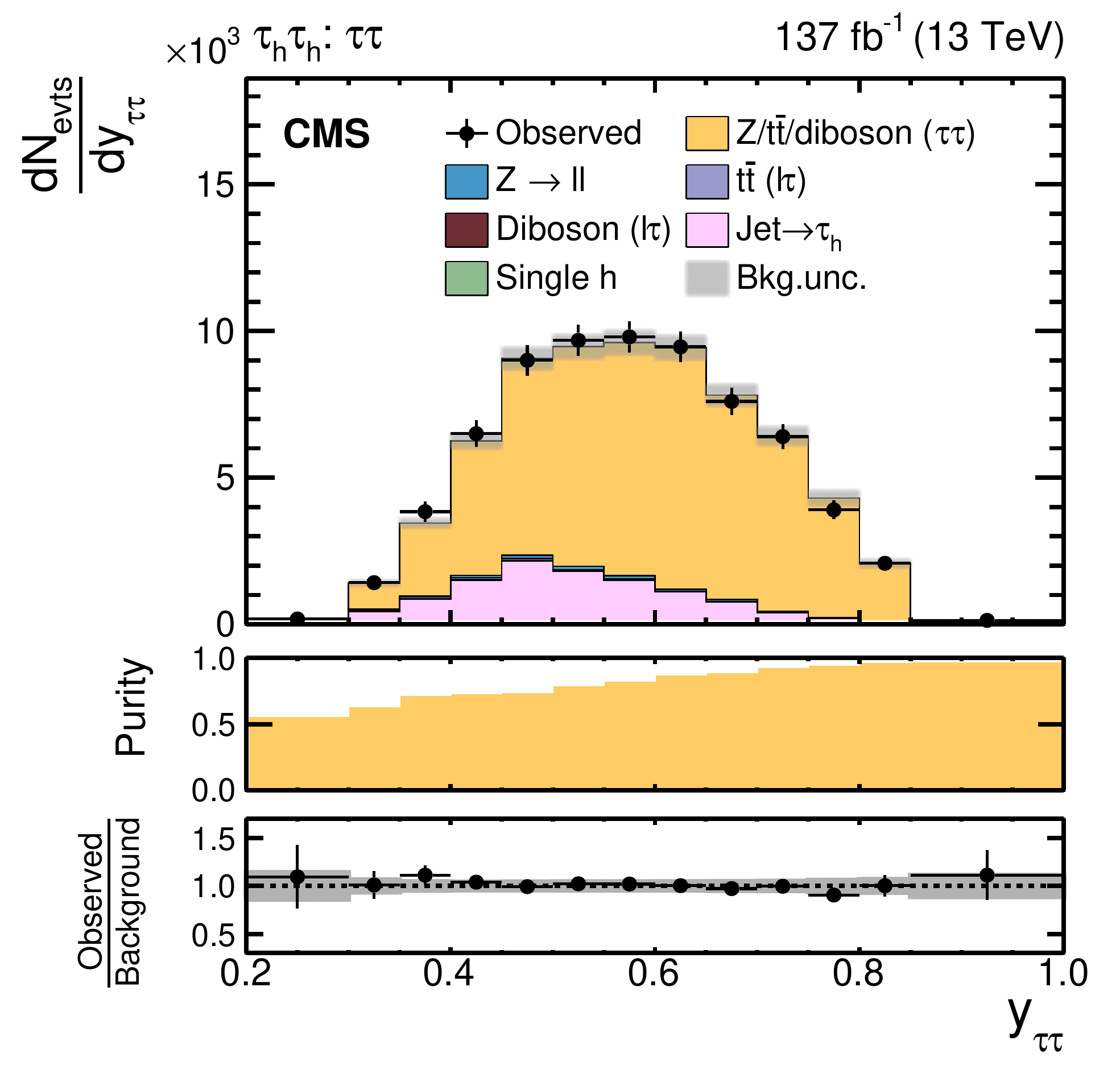}
  \includegraphics[width=0.45\textwidth]{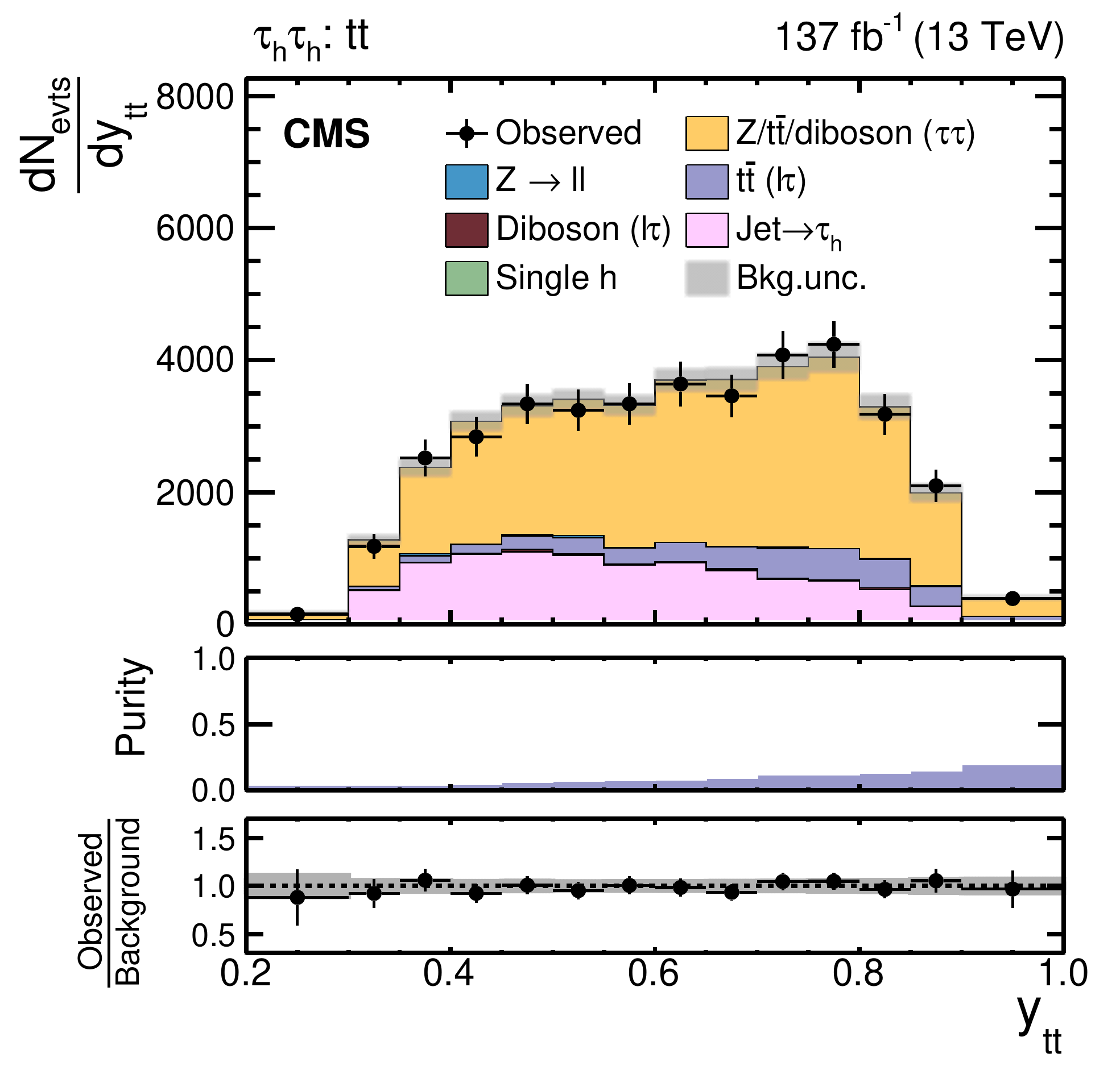}
  \includegraphics[width=0.45\textwidth]{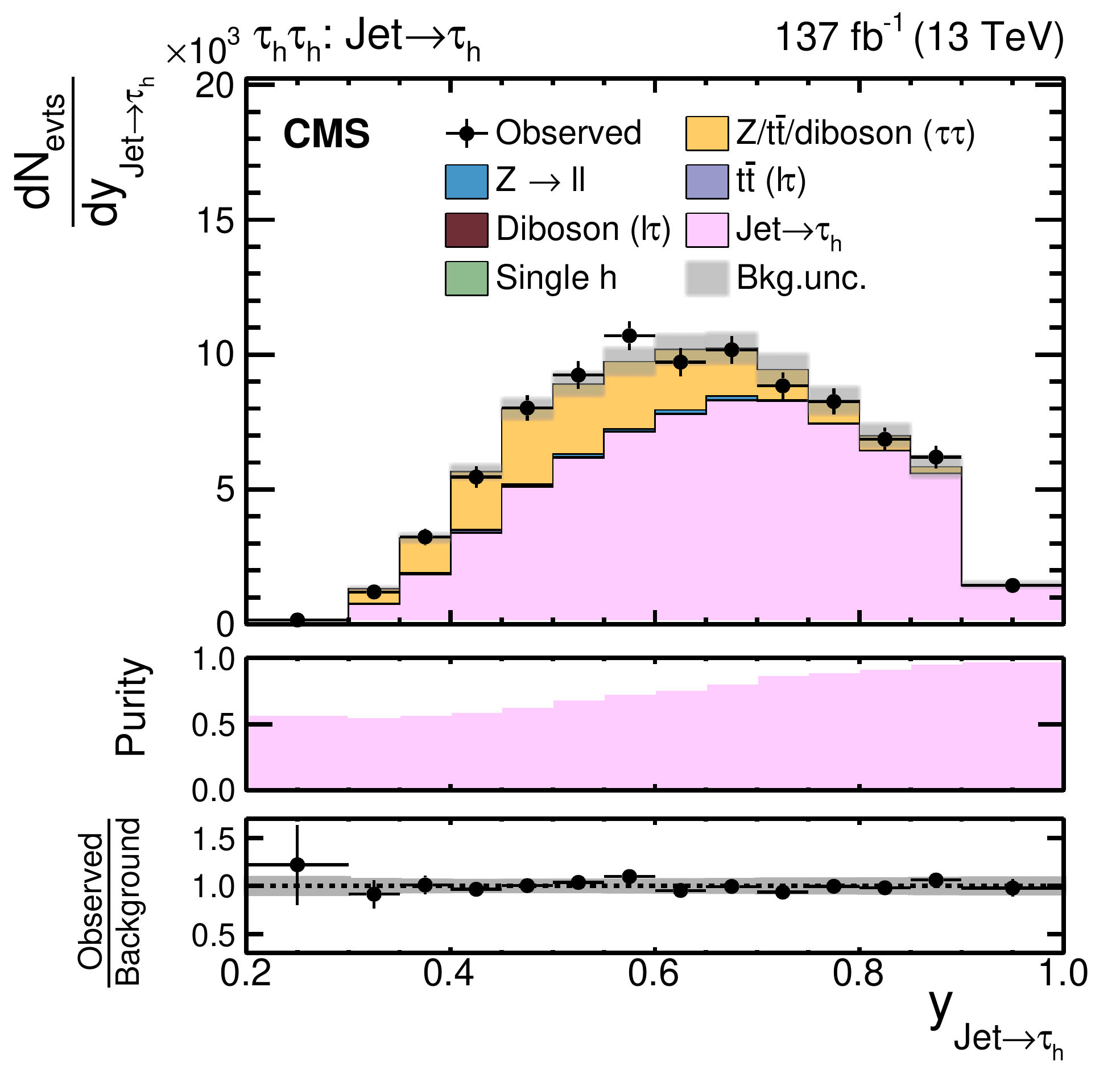}
  \includegraphics[width=0.45\textwidth]{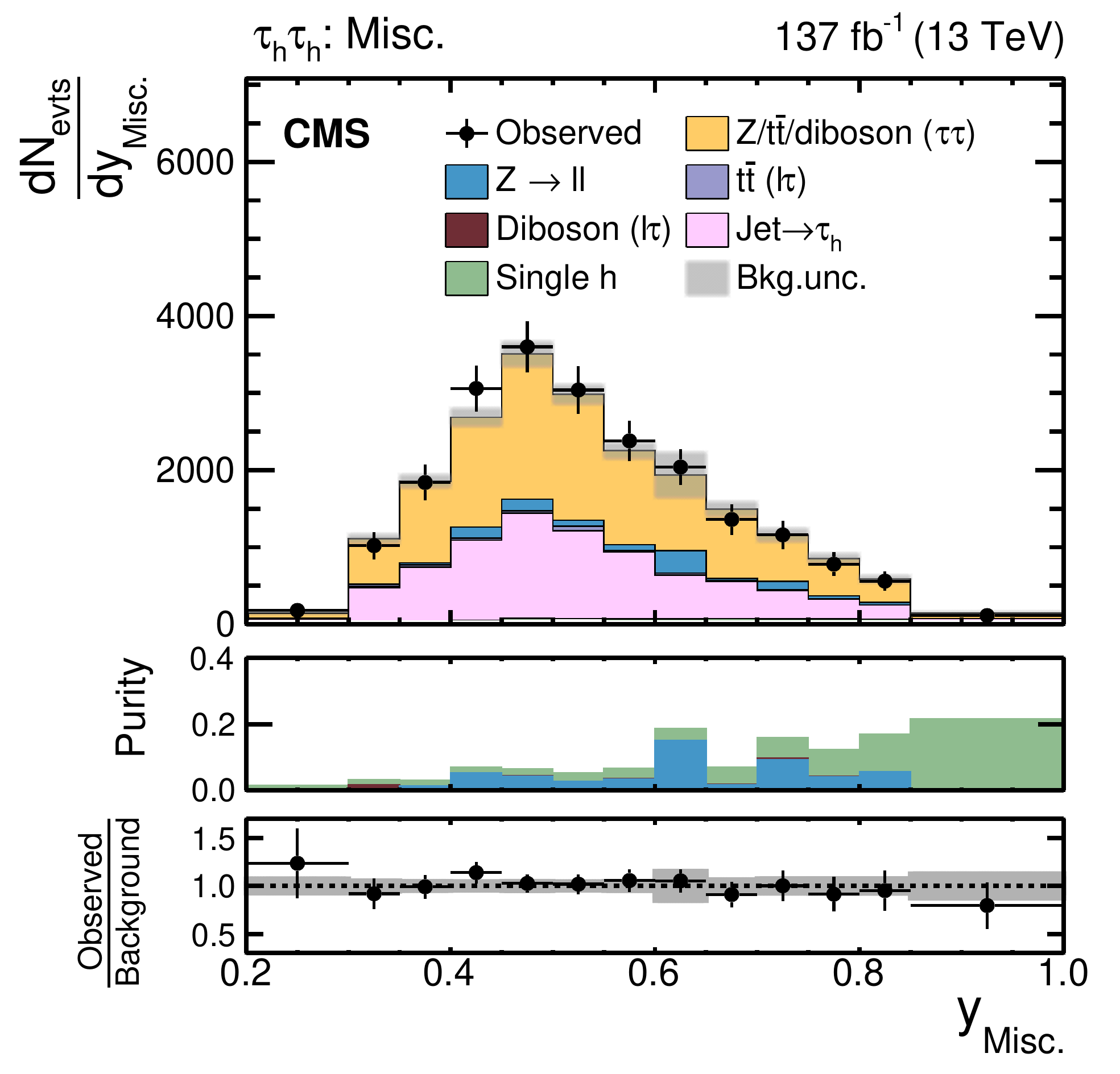}
  \includegraphics[width=0.45\textwidth]{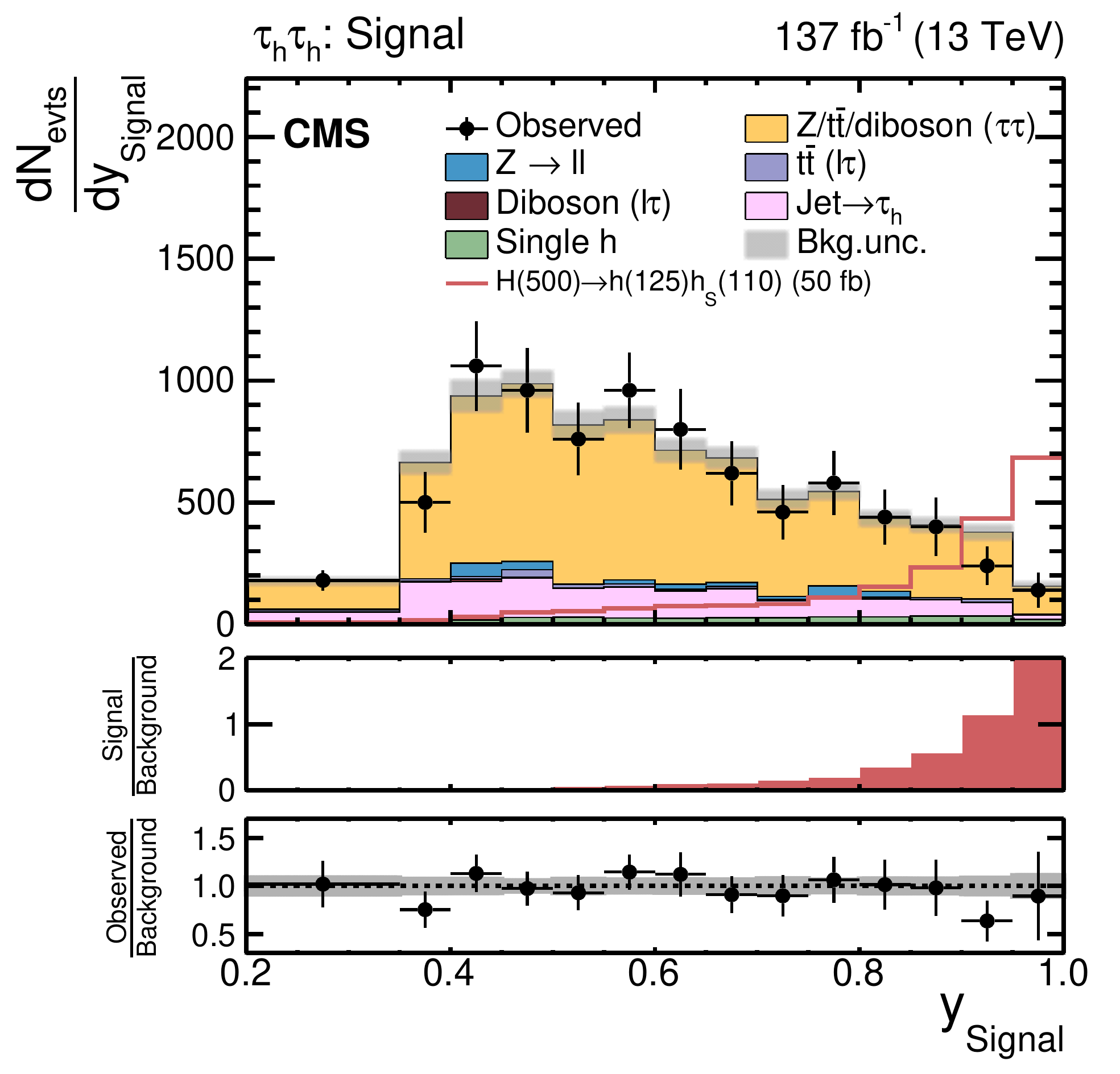}
  \hspace*{0.45\textwidth}
  \caption{
    Event categories after NN classification based on a training for $\mH=
    500\GeV$ and $100\leq \mhs<150\GeV$ in the $\tautau$ final state. Shown 
    are the (upper left) $\Pgt\Pgt$, (upper right) \ttc, (middle left) $\jettau$, 
    (middle right) misc, and (lower left) signal categories. For these figures 
    the data sets of all years have been combined. The uncertainty bands correspond 
    to the combination of statistical and systematic uncertainties after the fit 
    to the signal plus background hypothesis for $\mH=500\GeV$ and $\mhs
    =110\GeV$. 
  }
  \label{fig:postfit_tt}
\end{figure}

Sets of input distributions based on the NN classification for $\mH=500\GeV$ 
and $100\leq \mhs<150\GeV$ in the $\mutau$, $\etau$, and $\tautau$ final states 
are shown in Figs.~\ref{fig:postfit_mt}--\ref{fig:postfit_tt}. For these figures, 
the data from all three years of data-taking have been combined. To retain the 
shape of the distributions of the $y_{i}$ in each category, the histogram bins 
have been divided by their widths, in the upper panels of each figure. As a 
Bayesian probability estimate the values of $y_{i}$ range from $0.2$ to $1.0$. 
The lower bound is given by the constraint that each event has to be associated 
to one of the five event categories. In each event category, the targeted processes 
are expected to have increasing purity with increasing values of $y_{i}$. The 
points with error bars correspond to the data and the stacked filled histograms 
to the expectation from the background model. For the signal categories, the 
expectation for a signal with $\xsecDotBR=200$ or $50\unit{fb}$, depending on the 
$\Pgt\Pgt$ final state, is also shown by a red line.

In the middle panels for all background categories, the purity estimated for the 
background template of each corresponding event category is shown. For the signal 
categories, the ratio of the indicated signal divided by the sum of all backgrounds 
is shown. In the lower panels of each figure, the observed numbers of events divided 
by the numbers of events expected from the background model are shown for each 
bin. 

No signal-like excess is observed in any of the investigated mass combinations 
and $95\%$ confidence level ($\CL$) upper limits on the $\xsecDotBR$ of a 
potential signal are set following the modified frequentist approach as described 
in Refs.~\cite{Junk:1999kv,Read:2002hq}, using the same definition of the profile 
likelihood test statistic as defined in Refs.~\cite{CMS-NOTE-2011-005,Chatrchyan:2012tx}:
\begin{linenomath}
  \begin{equation}
    q_{\mu} = -2\ln
    \left(
    \frac{\mathcal{L}(\left.\{k_{i}\}\right|\mu\,S_{i}(\mH,\,\mhs,\,
    \{\hat{\theta}_{j,\mu}\})+
    B_{i}(\{\hat{\theta}_{j,\mu}\}))}
    {\mathcal{L}(\left.\{k_{i}\}\right|\hat{\mu}\,S_{i}(\mH,\,\mhs,\,
    \{\hat{\theta}_{j,\hat{\mu}}\})+
    B_{i}(\{\hat{\theta}_{j,\hat{\mu}}\}))}
    \right) , \;\; 0\leq \hat{\mu} \leq \mu.
    \label{eq:likelihood-ratio}
  \end{equation}
\end{linenomath}
In Eq.~(\ref{eq:likelihood-ratio}), $\hat{\mu}$, $\hat{\theta}_{j,\mu}$, and 
$\hat{\theta}_{j,\hat{\mu}}$ indicate the maximum likelihood estimates of the 
corresponding parameters from the fit to the data and the index of $q_{\mu}$ 
indicates that the fit to the data has been performed for a fixed value of $\mu$. 
In the large number limit, the distribution of $q_{\mu}$ can be approximated by 
analytic functions, from which the median and the uncertainty contours can be 
obtained as described in Ref.~\cite{Cowan:2010js}. 

\begin{figure}
  \centering
  \includegraphics[width=0.8\textwidth]{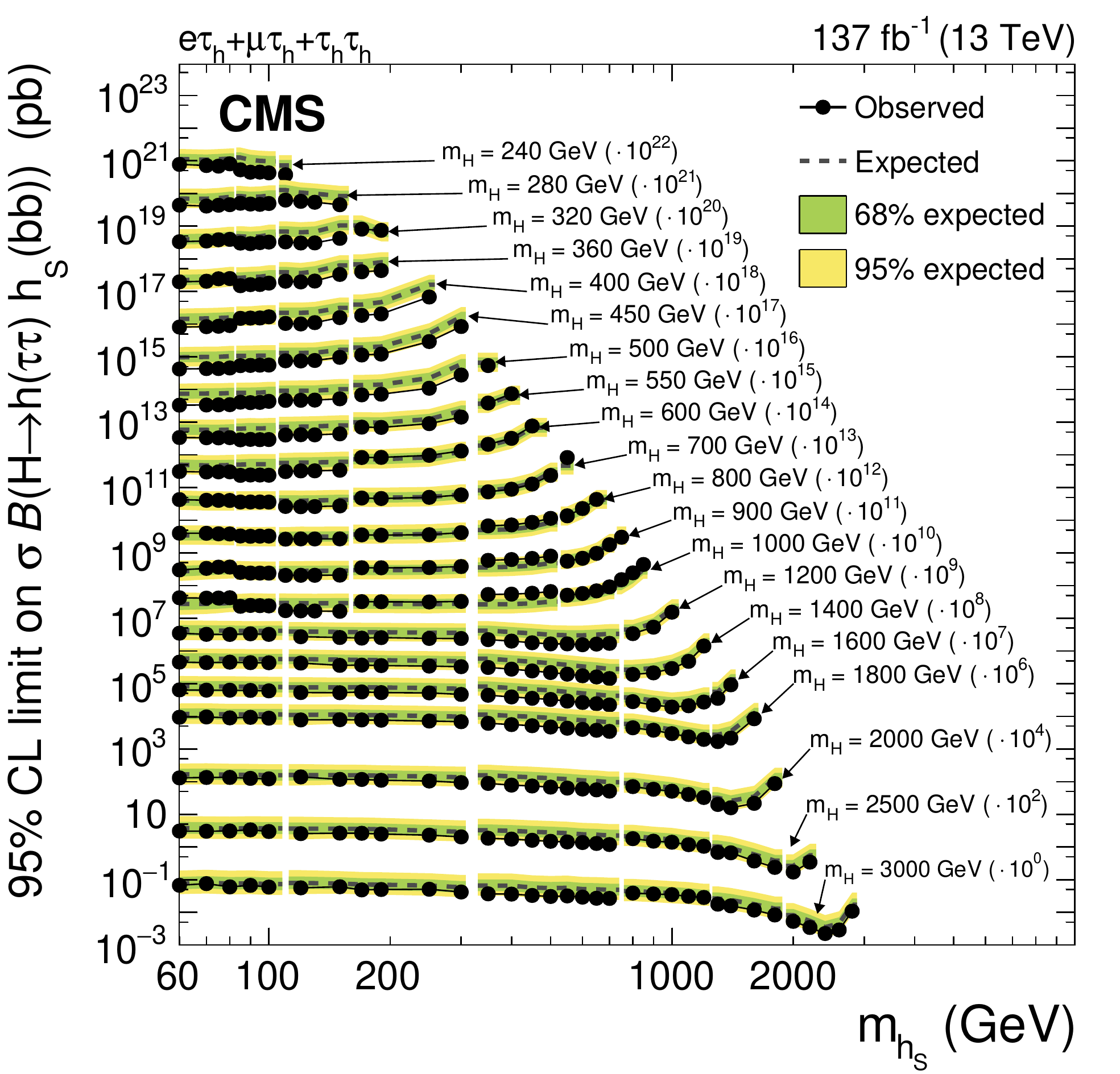}
  \caption{
    Expected and observed $95\%$ $\CL$ upper limits on $\xsecDotBR$ for all tested 
    values of $\mH$ and $\mhs$. The limits for each corresponding mass value have 
    been scaled by orders of ten as indicated in the annotations. Groups of hypothesis 
    tests based on the same NN trainings for classification are indicated by discontinuities 
    in the limits, which are linearly connected otherwise to improve the visibility 
    of common trends.
  }
  \label{fig:1d_limits}
\end{figure}

The observed and expected limits as a function of the tested values of $\mhs$ 
in a mass range from $240\leq \mH\leq3000\GeV$ and for the combination of all 
$\Pgt\Pgt$ final states and the analyzed data from all years are shown in 
Fig.~\ref{fig:1d_limits}. The observed limits are given by the black points. The 
expected median values in the absence of signal are indicated by the dashed black 
line with the central $68$ and $95\%$ expected quantiles for the upper limit given 
by the green and yellow bands. They range from $125\unit{fb}$ for $m_{\PH}=240
\GeV$ and $\mhs=85\GeV$ to $2.7\unit{fb}$ for $\mH=1000\GeV$ and $\mhs=350\GeV$ 
with a roughly flattening progression beyond. These limits are model independent. 
Since the analysis is not able to distinguish between scalar and pseudoscalar 
Higgs bosons, the limits are equally applicable to both cases. Residual differences 
on the detector acceptance for a scalar or pseudoscalar $\Phs$ are expected 
to be small and well covered by the theoretical acceptance uncertainties discussed 
in Section~\ref{sec:uncertainties}.

It should be noted that neighboring points in $\mhs$ differ only slightly in the 
kinematic properties of the tested signal hypotheses. Groups of hypothesis tests 
based on the same NN trainings for classification are indicated by discontinuities 
in the limits, which are linearly connected otherwise to improve the visibility 
of common trends.   

\begin{figure}[h!]
  \centering
  \includegraphics[width=0.8\linewidth]{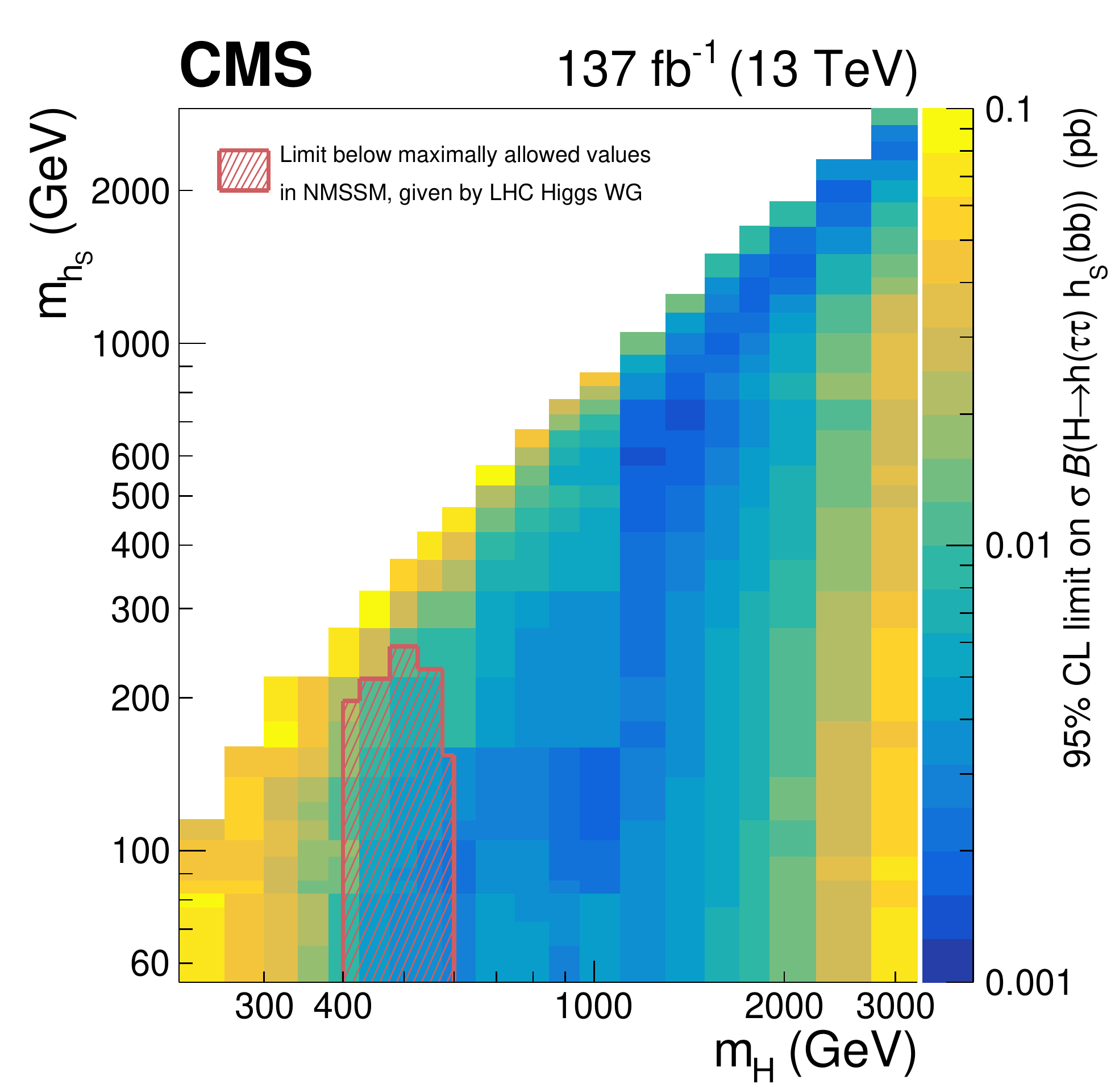}
  \caption{
    Summary of the observed limits on $\xsecDotBR$ for all tested pairs of $\mH$ 
    and $\mhs$, as shown in Fig.~\ref{fig:1d_limits}. The limits are given by the 
    color code of the figure. The region in the plane spanned by $\mH$ and $\mhs$ 
    where the observed limits fall below the maximally allowed values on $\xsecDotBR$ 
    in the context of the NMSSM, as provided by the LHC Higgs Working Group, are 
    indicated by a red hatched area.
  }
  \label{fig:mass_constraint}
\end{figure}

A summary of the observed limits for all tested pairs of $\mH$ and $\mhs$ is shown 
in Fig.~\ref{fig:mass_constraint}, where the limits are given by the color code 
of the figure.

Maximally allowed values for $\xsecDotBR$ in the context of the NMSSM for given 
pairs $\mH$ and $\mhs$, have been provided by the LHC Higgs Working Group, using 
the codes \textsc{NMSSMTools 5.5.0}~\cite{Ellwanger:2005dv} and 
\textsc{NMSSMCALC}~\cite{Baglio:2013iia}, incorporating experimental constraints 
from measurements of the $\Ph$ properties, SUSY searches, B-meson physics and dark 
matter searches. The region in the plane spanned by $\mH$ and $\mhs$ where the 
observed limits fall below these maximally allowed values on $\xsecDotBR$ are 
indicated by a read hatched area. It corresponds to $400\leq\mH\lesssim600\GeV$ 
and $60\leq\mhs\lesssim200\GeV$. For $m(\PH)=450\GeV$ and $60\leq\mhs\leq80\GeV$ 
the observed limits are five times smaller than the maximally allowed values for 
$\xsecDotBR$. Tabulated results of this analysis are available in the HepData 
database~\cite{hepdata}. 

\section{Summary}
\label{sec:summary}

A search for a heavy Higgs boson $\PH$ decaying into the observed Higgs boson 
$\Ph$ with a mass of $125\GeV$ and another Higgs boson $\Phs$ has been presented. 
The $\Ph$ and $\Phs$ bosons are required to decay into a pair of tau leptons and 
a pair of $\PQb$ quarks, respectively. The search uses a sample of proton-proton 
collisions collected with the CMS detector at a center-of-mass energy of $13\TeV$, 
corresponding to an integrated luminosity of $137\fbinv$. Mass ranges of $240$--$
3000\GeV$ for $\mH$ and $60$--$2800\GeV$ for $\mhs$ are explored in the search. No 
signal has been observed. Model independent $95\%$ confidence level upper limits 
on the product of the production cross section and the branching fractions of the 
signal process are set with a sensitivity ranging from $125\unit{fb}$ (for $\mH=
240\GeV$) to $2.7\unit{fb}$ (for $\mH=1000\GeV$). These limits have been compared 
to maximally allowed products of the production cross section and the branching 
fractions of the signal process in the next-to-minimal supersymmetric extension 
of the standard model. This is the first search for such a process carried out at 
the LHC.

\clearpage

\begin{acknowledgments}
  We congratulate our colleagues in the CERN accelerator departments for the excellent performance of the LHC and thank the technical and administrative staffs at CERN and at other CMS institutes for their contributions to the success of the CMS effort. In addition, we gratefully acknowledge the computing centers and personnel of the Worldwide LHC Computing Grid and other centers for delivering so effectively the computing infrastructure essential to our analyses. Finally, we acknowledge the enduring support for the construction and operation of the LHC, the CMS detector, and the supporting computing infrastructure provided by the following funding agencies: BMBWF and FWF (Austria); FNRS and FWO (Belgium); CNPq, CAPES, FAPERJ, FAPERGS, and FAPESP (Brazil); MES (Bulgaria); CERN; CAS, MoST, and NSFC (China); MINCIENCIAS (Colombia); MSES and CSF (Croatia); RIF (Cyprus); SENESCYT (Ecuador); MoER, ERC PUT and ERDF (Estonia); Academy of Finland, MEC, and HIP (Finland); CEA and CNRS/IN2P3 (France); BMBF, DFG, and HGF (Germany); GSRT (Greece); NKFIA (Hungary); DAE and DST (India); IPM (Iran); SFI (Ireland); INFN (Italy); MSIP and NRF (Republic of Korea); MES (Latvia); LAS (Lithuania); MOE and UM (Malaysia); BUAP, CINVESTAV, CONACYT, LNS, SEP, and UASLP-FAI (Mexico); MOS (Montenegro); MBIE (New Zealand); PAEC (Pakistan); MSHE and NSC (Poland); FCT (Portugal); JINR (Dubna); MON, RosAtom, RAS, RFBR, and NRC KI (Russia); MESTD (Serbia); SEIDI, CPAN, PCTI, and FEDER (Spain); MOSTR (Sri Lanka); Swiss Funding Agencies (Switzerland); MST (Taipei); ThEPCenter, IPST, STAR, and NSTDA (Thailand); TUBITAK and TAEK (Turkey); NASU (Ukraine); STFC (United Kingdom); DOE and NSF (USA).

  \hyphenation{Rachada-pisek} Individuals have received support from the Marie-Curie program and the European Research Council and Horizon 2020 Grant, contract Nos.\ 675440, 724704, 752730, 765710 and 824093 (European Union); the Leventis Foundation; the Alfred P.\ Sloan Foundation; the Alexander von Humboldt Foundation; the Belgian Federal Science Policy Office; the Fonds pour la Formation \`a la Recherche dans l'Industrie et dans l'Agriculture (FRIA-Belgium); the Agentschap voor Innovatie door Wetenschap en Technologie (IWT-Belgium); the F.R.S.-FNRS and FWO (Belgium) under the ``Excellence of Science -- EOS" -- be.h project n.\ 30820817; the Beijing Municipal Science \& Technology Commission, No. Z191100007219010; the Ministry of Education, Youth and Sports (MEYS) of the Czech Republic; the Deutsche Forschungsgemeinschaft (DFG), under Germany's Excellence Strategy -- EXC 2121 ``Quantum Universe" -- 390833306, and under project number 400140256 - GRK2497; the Lend\"ulet (``Momentum") Program and the J\'anos Bolyai Research Scholarship of the Hungarian Academy of Sciences, the New National Excellence Program \'UNKP, the NKFIA research grants 123842, 123959, 124845, 124850, 125105, 128713, 128786, and 129058 (Hungary); the Council of Science and Industrial Research, India; the Latvian Council of Science; the Ministry of Science and Higher Education and the National Science Center, contracts Opus 2014/15/B/ST2/03998 and 2015/19/B/ST2/02861 (Poland); the National Priorities Research Program by Qatar National Research Fund; the Ministry of Science and Higher Education, project no. 0723-2020-0041 (Russia); the Programa Estatal de Fomento de la Investigaci{\'o}n Cient{\'i}fica y T{\'e}cnica de Excelencia Mar\'{\i}a de Maeztu, grant MDM-2015-0509 and the Programa Severo Ochoa del Principado de Asturias; the Thalis and Aristeia programs cofinanced by EU-ESF and the Greek NSRF; the Rachadapisek Sompot Fund for Postdoctoral Fellowship, Chulalongkorn University and the Chulalongkorn Academic into Its 2nd Century Project Advancement Project (Thailand); the Kavli Foundation; the Nvidia Corporation; the SuperMicro Corporation; the Welch Foundation, contract C-1845; and the Weston Havens Foundation (USA).
  \end{acknowledgments}

\clearpage 

\bibliography{auto_generated}

\cleardoublepage \appendix\section{The CMS Collaboration \label{app:collab}}\begin{sloppypar}\hyphenpenalty=5000\widowpenalty=500\clubpenalty=5000\input{HIG-20-014-authorlist.tex}\end{sloppypar}
%%% END EDITABLE REGION %%%
% skeleton_end
\end{document}

%% file: HIG-20-014-authorlist.tex
\vskip\cmsinstskip
\textbf{Yerevan Physics Institute, Yerevan, Armenia}\\*[0pt]
A.~Tumasyan
\vskip\cmsinstskip
\textbf{Institut f\"{u}r Hochenergiephysik, Wien, Austria}\\*[0pt]
W.~Adam, J.W.~Andrejkovic, T.~Bergauer, S.~Chatterjee, M.~Dragicevic, A.~Escalante~Del~Valle, R.~Fr\"{u}hwirth\cmsAuthorMark{1}, M.~Jeitler\cmsAuthorMark{1}, N.~Krammer, L.~Lechner, D.~Liko, I.~Mikulec, P.~Paulitsch, F.M.~Pitters, J.~Schieck\cmsAuthorMark{1}, R.~Sch\"{o}fbeck, M.~Spanring, S.~Templ, W.~Waltenberger, C.-E.~Wulz\cmsAuthorMark{1}
% ...
\vskip\cmsinstskip
% ...
1:  Also at TU Wien, Wien, Austria\\
% ...
